%% file: arxiv.tex
\pdfoutput=1
\RequirePackage{lineno}
\newif\ifpreprint%
\preprintfalse%
\ifpreprint%
	\documentclass[preprint,english,aps,prb,floatfix,amssymb,superscriptaddress,a4paper]{revtex4}
\else%
	\documentclass[twocolumn,english,aps,prb,floatfix,amssymb,superscriptaddress,a4paper]{revtex4}
\fi%
\usepackage{amsmath}
\usepackage{siunitx}
\usepackage{dsfont}
\usepackage{color}
\usepackage{amssymb}
\usepackage{graphicx}
\usepackage{braket}
\usepackage{units}
\usepackage{chemformula}
\usepackage{amssymb}
\usepackage{mathtools}
\usepackage{physics}
\usepackage{braket}
\usepackage{bibunits}

\usepackage{hyperref}
\hypersetup{
  colorlinks = true,
  hypertexnames=true
}

\renewcommand{\theta}{\vartheta}
\renewcommand{\phi}{\varphi}

\newcommand{\RN}[1]{%
  \textup{\uppercase\expandafter{\romannumeral#1}}%
}

\begin{document}
\ifpreprint%
	\linenumbers%
\fi%

\title{Experimental characterization of fragile topology in an acoustic metamaterial}

\date{\today}

\author{Valerio Peri}
\affiliation{Institute for Theoretical Physics, ETH Zurich, 8093 Z\"urich, Switzerland}
\author{Zhi-Da Song}
\affiliation{Department of Physics, Princeton University, Princeton, New Jersey 08544, USA}
\author{Marc Serra-Garcia}
\affiliation{Institute for Theoretical Physics, ETH Zurich, 8093 Z\"urich, Switzerland}
\author{Pascal Engeler}
\affiliation{Institute for Theoretical Physics, ETH Zurich, 8093 Z\"urich, Switzerland}
\author{Raquel Queiroz}
\affiliation{Department of Condensed Matter Physics, Weizmann Institute of Science,Rehovot 76100, Israel}
\author{Xueqin Huang}
\affiliation{School of Physics and Optoelectronics, South China University of Technology, Guangzhou, Guangdong 510640, China}
\author{Weiyin Deng}
\affiliation{School of Physics and Optoelectronics, South China University of Technology, Guangzhou, Guangdong 510640, China}
\author{Zhengyou Liu}
\affiliation{Key Laboratory of Artificial Micro- and Nanostructures of Ministry of Education and School of Physics and Technology, Wuhan University, Wuhan 430072, China}
\affiliation{Institute for Advanced Studies, Wuhan University, Wuhan 430072, China}\
\author{B. Andrei Bernevig}
\affiliation{Department of Physics, Princeton University, Princeton, New Jersey 08544, USA}
\affiliation{Physics Department, Freie Universit\"at Berlin, Arnimallee 14, 14195 Berlin, Germany}
\affiliation{Max Planck Institute of Microstructure Physics, 06120 Halle, Germany}
\author{Sebastian D. Huber}
\affiliation{Institute for Theoretical Physics, ETH Zurich, 8093 Z\"urich, Switzerland}

\let\oldaddcontentsline\addcontentsline
\renewcommand{\addcontentsline}[3]{}
\begin{bibunit}[naturemag]
\input{manuscript_bare}

\input{bu1.bbl}
\end{bibunit}
\let\addcontentsline\oldaddcontentsline

	\end{document}


\thispagestyle{empty}
\begin{center}
  \textbf{\large Supplementary material: Experimental characterization of fragile topology in an acoustic metamaterial}\\[.2cm]
  Valerio Peri,$^{1}$ Zhi-Da Song,$^{2}$ Marc Serra-Garcia,$^{1}$ Pascal Engeler,$^{2}$ Raquel Queiroz,$^{3}$ Xueqin \\ Huang,$^{4}$ Weiyin Deng,$^{4}$ Zhengyou Liu,$^{5,6}$, B. Andrei Bernevig,$^{2,7,8}$ and Sebastian D. Huber$^{1}$ \\[.1cm]
  {\itshape ${}^1$Institute for Theoretical Physics, ETH Zurich, 8093 Z\"urich, Switzerland\\
  ${}^2$Department of Physics, Princeton University, Princeton, New Jersey 08544, USA\\
  ${}^3$Department of Condensed Matter Physics, Weizmann Institute of Science, Rehovot 76100, Israel\\
   ${}^4$School of Physics and Optoelectronics, South China University of Technology, Guangzhou, Guangdong 510640, China\\
    ${}^5$Key Laboratory of Artificial Micro- and Nanostructures of Ministry of Education and School of Physics and Technology, Wuhan University, Wuhan 430072, China\\
	 ${}^6$Institute for Advanced Studies, Wuhan University, Wuhan 430072, China\\
	  ${}^7$Physics Department, Freie Universit\"at Berlin, Arnimallee 14, 14195 Berlin, Germany\\
	   ${}^8$Max Planck Institute of Microstructure Physics, 06120 Halle, Germany\\}
(Dated: \today)\\[1cm]
\end{center}
 \input{supplementary_bare}

%% file: manuscript_bare.tex

\begin{abstract} 
Symmetries crucially underlie the classification of topological phases of matter. Most materials, both natural as well as architectured, possess crystalline symmetries. Recent theoretical works unveiled that these crystalline symmetries can stabilize fragile Bloch bands that challenge our very notion of topology: while answering to the most basic definition of topology, one can trivialize these bands through the addition of trivial Bloch bands. Here, we fully characterize the symmetry properties of the response of an acoustic metamaterial to establish the fragile nature of the low-lying Bloch bands. Additionally, we present a spectral signature in the form of spectral flow under twisted boundary conditions. 
\end{abstract} 
%

\maketitle
While topological properties of phases of matter seem to be an omnipresent theme in contemporary condensed matter research, there is no unique defining property of what a ``topological'' system is.\cite{Wen19} For strongly interacting phases, one might use as a definition the existence of fractionalized excitations or the presence of long-range entanglement.\cite{Wen91a} For non-interacting systems, bulk-boundary correspondences can often be captured by topological indices such as Chern or winding numbers.\cite{Chiu15} Despite the vast differences between the various instances of topological matter, all these phases have one common denominator: one cannot smoothly transform the system to an ``atomic limit'' of disconnected elementary blocks separated in space.

\begin{figure*}[bth!]
\includegraphics{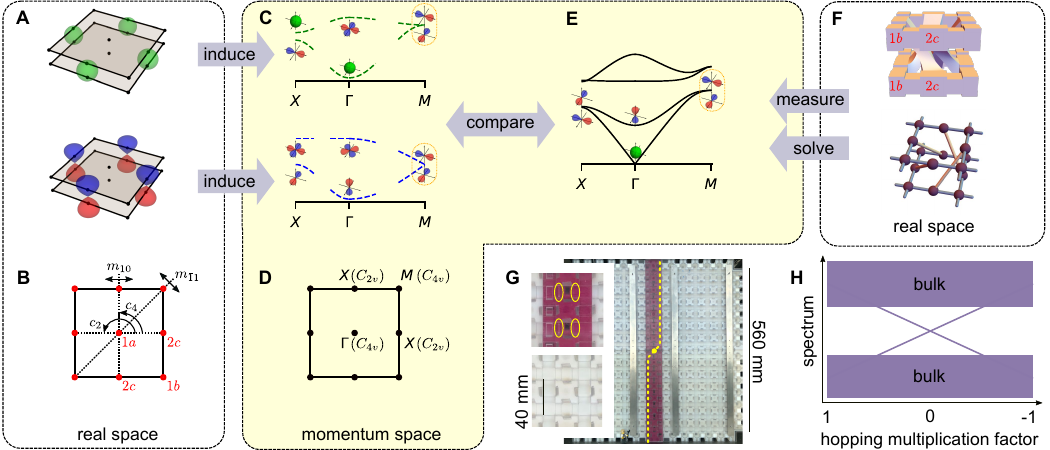}
\caption{{\bf Atomic limit and topological bands.} {\bf A} Localized orbitals at the $2c$ maximal Wyckoff positions in the unit cell of a $p4mm$--symmetric system: $A_1$ orbitals in the upper panel and $A_2$  orbitals in the lower one (see Fig.~\ref{fig:dispersion}.B and Table S3 \cite{si}). {\bf B} Schematic representation of the wallpaper group $p4mm$. In red, the locations and labels of the maximal Wyckoff positions. Dotted black lines indicate the relevant mirror planes with respective labels, while solid lines show the action of $c_4$ and $c_2$ symmetry operators. {\bf C} Sketch of the bands at high-symmetry points induced by the localized orbitals of {\bf A}.\cite{Bradlyn:2017,Bilbao,topoChem} The drawings show example orbitals that transform according to the realized irreducible representations. {\bf D} Labels and locations of the high-symmetry points in the Brillouin zone for the $p4mm$ wallpaper group. In parenthesis, the little group realized at each high-symmetry point. {\bf E} Bands obtained from finite-element simulations of our acoustic crystal. The irreducible representations at high-symmetry points are represented by example orbitals. {\bf F} The upper panel shows a rendering of the air structure of the acoustic crystal unit cell with $p4mm$ symmetry. In red, the labels of the maximal Wyckoff positions. In the lower panel, a lattice representation of the acoustic structure. {\bf G} Photo of the experimental sample with the soft cut indicated by the yellow dashed line. The upper panel of the inset shows a detail of the obstructions realizing the cut, the lower panel a zoom-in of the unit cell. {\bf H} Schematic of the flow induced between fragile bands under twisted boundary conditions.}
\label{fig:setup}
\end{figure*}
For the classification of such topological systems, symmetries play an essential role. The path of smooth transformations to an atomic limit can be obstructed by symmetry constraints.\cite{Chen11a} Prime examples are the table of non-interacting topological insulators,\cite{Kitaev09, Altland97,Bernevig06a,Konig07} or the Affleck-Kennedy-Lieb-Tasaki \cite{Affleck87} spin-1 state as a member of the family of symmetry protected interacting phases in one dimension.\cite{Chen11a} Lately, crystalline symmetries have been identified as an extremely rich source of topological band structures for electrons.\cite{Bradlyn:2017,Hughes:2011,Fu11,Po17,Song18} In fact, this applies not only to electronic bands but to any periodic linear system, both quantum and classical. 

It has been realized that crystalline topological insulators can be divided into two main classes based on their stability under the addition of other bands.\cite{Po:2018,Bradlyn:2017} Stable topological insulators can only be trivialized by another set of topological bands, and can be classified using K-theory.\cite{Kitaev09} Contrarily, there are topological bands that can be trivialized by a set of bands arising from an atomic limit. In this case, our conventional notion of topological robustness is challenged and one needs to introduce the idea of {\em fragile topology}.\cite{Po:2018,Bouhon:2019,Song:2019, Alexandradinata:2019, Hwang:2019, Liu:2019, Wieder:2018, Bradlyn:2019} To understand how this fragility arises, one needs to classify the bands emerging from the 230 crystalline space groups.

Recent approaches to this challenge attempt this classification by ``inverting'' the idea of the atomic limit:\cite{Bradlyn:2017,Bradlyn:2018} starting from a set of isolated orbitals at high-symmetry points in {\em real space}, one constructs all possible bands that can be induced from these orbitals. These are called elementary band representations, or EBRs.\cite{Bradlyn:2017,Cano:2018, Zak:1982} Such an EBR is characterized by the irreducible representations of the symmetry realized at high-symmetry locations in the Brillouin zone.  Once all such EBRs are constructed, one can compare them to a concrete set of bands obtained from an experiment or numerical calculations. If the bands under investigation can be written as a combination of bands induced from isolated orbitals, one can perform the atomic limit by construction. In essence, the classification task is turned into a simple matching exercise between the possible EBRs and bands realized in a material.

In case this matching exercise fails, two distinct ways of how a set of bands can be topological arise. First, the physical bands might realize only a fraction of the representations of an EBR. In other words, the EBR induced from a given position splits into disconnected parts. Once we take only ``half'' the EBR, the atomic limit is obstructed and we deal with a topological system.\cite{Bradlyn:2017,Cano:2018a} This corresponds to the familiar case of stable topology.

In the second option, the multiplicities of the representations can be reproduced by combining different EBRs with {\em positive} and {\em negative} coefficients. This means the physical bands lack a number of high-symmetry-point representations to be compatible with an atomic limit. However, the lacking representations correspond to an EBR. In other words, an atomic limit is possible, if one adds the bands of this ``missing'' EBR. This matches the definition of {\em fragile topology} given above. 

Is fragile topology an exotic curiosity and, moreover, does it have any experimental signatures in spectral or transport measurements? The first question can be answered with an affirmative no. It turns out that fragile band-structures are abundant both in electronic materials,\cite{Song:2019} e.g., the flat bands of magic angle twisted bilayer graphene,\cite{Cao18,Ahn:2019,Po:2019,Zaletel:2019,Lian:2018} as well as for classical systems in photonics and phononics.\cite{Wang:2019a,Blanco-de-Paz:2019,Alexandradinata:2019} The second question was answered in a recent complementary work.\cite{Zhida} When the system with fragile bands is slowly disconnected into several parts while preserving some of the space group symmetries, spectral flow is occurring: a number of states determined by the topology of the involved bands is flowing through the bulk gap.\cite{Zhida} This work presents an experimental characterization of this spectral flow, thereby turning the abstract concept of fragile topology into a measurable effect.

\begin{figure*}[tbh]
\includegraphics{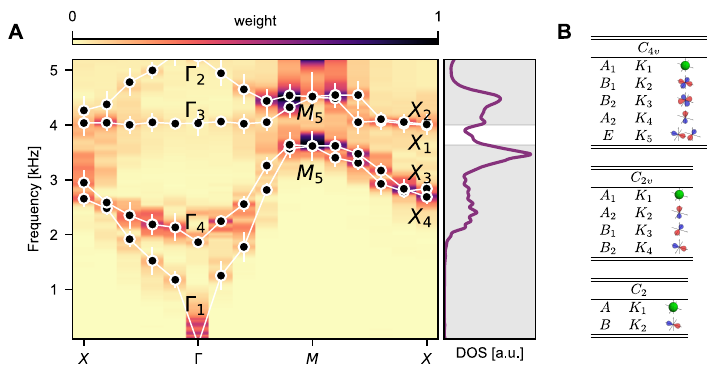}
\caption{{\bf Irreducible representations at high symmetry points.} {\bf A} Measured spectrum of the acoustic crystal along high symmetry lines. The fit to the local maxima has been overlaid at each point in momentum space and the vertical error bars are the full-width-half-maxima of the fitted Lorentzians. Labels indicate the irreducible representations of the little groups $G_K$ realized at high symmetry points $K$ according to the names in {\bf B} ($G_\Gamma=G_M\cong C_{4v}$ and $G_X\cong C_{2v}$ ). In the right panel, we show the integrated density of states. The frequency range of the bulk bands is shaded in gray. {\bf B} Tables for the irreducible representations of $C_{4v}$ and its relevant subgroups: $C_{2v}$ and $C_2$. The first column gives the standard name according to Ref.[\onlinecite{Cracknell72}], the second column the label we give at the high-symmetry points in the Brillouin zone (e.g., $K_1\rightarrow\Gamma_1$ at the $\Gamma$-point), the third column depicts an example orbital in the respective irreducible representation.}
\label{fig:dispersion}
\end{figure*}

We constructed an acoustic sample in the wallpaper group $p4mm$ to experimentally establish the spectral signatures of fragile bands. The symmetries and the high-symmetry points (maximal Wyckoff positions) in the unit cell of $p4mm$ are explained in Fig.~\ref{fig:setup}.B. Our sample is made of two layers, where the acoustic cavities reside at the maximal Wyckoff positions $1b$ and $2c$. Chiral coupling channels connect the $2c$--cavities in different layers, see Fig.~\ref{fig:setup}.F. 

When considering the low-frequency Bloch bands, one can expect the nodeless modes of the $2c$ cavities to be relevant as they have a larger volume. Specifically, we presume the orbitals $A_1$ and $A_2$ at $2c$ to induce the lowest four bands, cf. Fig.~\ref{fig:setup}.A. Comparing the bands induced from these orbitals in Fig.~\ref{fig:setup}.C to the result of the finite-element simulation of the acoustic field\cite{si} in Fig.~\ref{fig:setup}.E, one observes that the representation at the high-symmetry points of the two lowest bands cannot be written by a combination of the EBRs induced from the expected orbitals. This points towards topological bands.

We now turn to the experimental validation. The acoustic crystal in Fig.~\ref{fig:setup}.G is fabricated by 3D-printing acoustically hard walls around the air-volume depicted in Fig.~\ref{fig:setup}.F. We excite the air at a fixed cavity in the middle of the sample via a speaker.\cite{si} By measuring at the center of each cavity, we obtain the Greens function $G_{\alpha,\beta}^{i,j}=\langle \psi^*_{\alpha}(i)\psi_{\beta}(j)\rangle$, where $\psi_\beta(j)$ denotes the acoustic field at the speaker in unit cell $j$ and Wyckoff position $\beta$, and likewise for the measured field $\psi_\alpha(i)$.\cite{si} We Fourier-transform the result to obtain the spectral information shown in Fig.~\ref{fig:dispersion}.A. Evaluating the relative weight and phase at different Wyckoff positions inside the unit cell allows us to extract the symmetry of the Bloch wave-functions \cite{si}, see Fig.~S4--S5. In Fig.~\ref{fig:dispersion}.A, we label the high-symmetry points according to the irreducible representation of the respective little-group.\cite{Elcoro:2017} The names and example orbitals of the representations are shown in Fig.~\ref{fig:dispersion}.B.

The measured symmetries confirm the expectations from the finite-element simulations. We find the following decomposition:\cite{Elcoro:2017}
\begin{align}
     \label{eqn:decomp}
     \text{Bands 1\&2}&:\quad (A_1)_{1b}\oplus (A_2)_{2c} \ominus (B_2)_{1a}, \\
     \text{Bands 3\&4}&:\quad(B_2)_{1a}\oplus (A_1)_{2c} \ominus (A_1)_{1b}.
\end{align}
This establishes that the lowest two sets of bands are fragile based on experimental data alone. The bands 1\&2 have the ``missing'' EBR induced from Wyckoff position $1a$ which does not host an acoustic cavity. In this case, fragility has a spectral consequence in the form of spectral flow. We quickly review how to establish this. Details can be found in the companion paper [\onlinecite{Zhida}].

\begin{figure*}[tbh]
\includegraphics{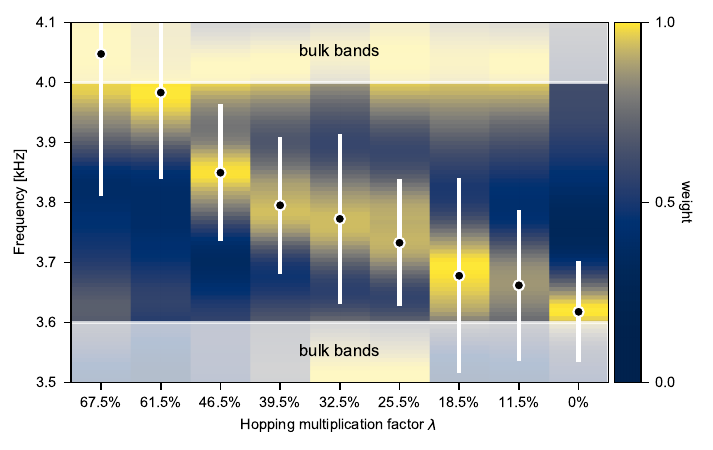}
\caption{{\bf Spectral flow between fragile bands.} Measured local density of states at the symmetry center of the acoustic crystal for different values $\lambda$ of the hopping multiplication factor across the cut. The overlaid dots indicate the fit to the local maxima and the vertical error bars correspond to the full-width-half-maxima of the fitted Lorentzians. The frequency range of the bulk bands has been shaded in gray. The data at each $\lambda$ has been individually normalized with respect to its maximum value.}
\label{fig:flow}
\end{figure*}

The simplest approach is given by a real-space picture.\cite{Zhida} One can characterize all states below the spectral gap of a finite sample by their transformation properties under the $C_2$ symmetry around the central $1a$ position. This allows us to write a real space index (RSI) \cite{Zhida}
\begin{equation}
	\delta = m(B)-m(A)
\end{equation}
which counts the imbalance between $C_2$--odd and $C_2$--even states, cf. Fig.~\ref{fig:dispersion}.B. One can now smoothly disconnect the sample in a $C_2$--symmetric way, cf. Fig.~\ref{fig:setup}.G, and reconnect it with the opposite hopping sign. The initial and final dynamics are related by a gauge transformation, where we multiply all degrees of freedom in one half of the sample by minus one. However, this gauge-transformation turns $C_2$--even into $C_2$--odd states. Hence, we expect $\delta$ states flowing through the gap during this cutting and reconnecting procedure. 

The remaining task is to bind this RSI $\delta$ to the decomposition in Eq. (\ref{eqn:decomp}). This can be achieved by an exhaustive analysis of all EBRs in a given space group, akin to the matching exercise to establish fragility.\cite{Zhida} For our situation we find \cite{si}
\begin{equation}
	\delta=1.
\end{equation}

How do we implement this soft cut in our sample? We insert obstructions of growing sizes into the channels along the yellow line in Fig.~\ref{fig:setup}.G.  Without using higher-order resonances in the connecting channels,\cite{Cummer16} we can only perform half of the twisting cycle: from the original sample ($\lambda=1$) to a total disconnect ($\lambda=0$), where $\lambda$ is the hopping multiplication factor across the cut. However, the spectrum at $\lambda$ and $-\lambda$ is related by a gauge transformation. Hence, the path $\lambda=0\to -1$ does not provide further information.\cite{si} 

Figure~\ref{fig:flow} shows the measured spectral flow. One observes how a state is flowing from the upper bulk bands to the lower bulk bands in the course of the cutting procedure. We obtain these results from the local density of states measured at the symmetry center of the sample. This strategy allows us to obtain a clean and robust spectral signature associated to fragile topology. Note that the opposite flow of a $C_2$--odd state from the lower bands to upper ones occur for $\lambda<0$ as ensured by the symmetry $\lambda\rightarrow -\lambda$.\cite{Zhida}

Through the results presented here, we achieved two goals. First, we established the presence of fragile bands through full band-tomography. Second, we demonstrate that fragile topology, a concept that challenges our understanding of topology can yield a clean and simple experimental observable. In the context of metamaterials design, such exquisite control of well-localized states is an important building block for mechanical logic and other wave-control applications.\cite{Pal19} Moreover, many classical topological metamaterials rely on crystalline symmetries.\cite{Wu:2015} This attributes fragility a more prominent role than anticipated.\cite{Wang:2019a,Blanco-de-Paz:2019,Alexandradinata:2019} Finally, the expectation that fragility plays a role in the reported strongly correlated superconductivity in twisted bi-layer graphene \cite{Cao18, Ahn:2019, Po:2019,Song:2019a, Zaletel:2019, Lian:2018} raises the natural question of how classical non-linearities are influenced by these intricate band effects. 

\bigskip
\noindent
{\bf Acknowledgements}: We thank Gianni Blatter, Titus Neupert and Vincenzo Vitelli for insightful discussions. {\bf Funding}: S. D. H., V. P., M. S. -G. and P. E. acknowledge support from the Swiss National Science Foundation, the NCCR QSIT, and the European Research Council under the Grant Agreement No. 771503 (TopMechMat). Z. S. and B. A. B. are supported by the Department of Energy Grant No. DE-SC0016239, the National Science Foundation EAGER Grant No. DMR1643312, Simons Investigator Grants No. 404513, ONRNo. N00014-14-1-0330, and NSF-MRSEC No. DMR-142051, the Packard Foundation, the Schmidt Fund for Innovative Research. B. A. B. is also supported by a Guggenheim Fellowship from the John Simon Guggenheim Memorial Foundation. {\bf Author contributions}: S.D.H., V.P., Z.S., R.Q. and B.A.B performed the theoretical part of this work. S.D.H., V.P., X.H., W.D. and Z.L. designed the samples. V.P., M.S. -G. and P.E. conducted the experiment. All authors contributed to the writing of the manuscript. {\bf Competing interests}: The authors declare no competing interests. {\bf Data and materials availability}: The data shown in this work are available at Ref.[\onlinecite{data}]. 

%% file: supplementary_bare.tex
\tableofcontents
\setcounter{page}{1}
\section{Materials and Methods}
\label{section:methods}
\subsection{Details of the sample design and fabrication.} 

\begin{figure}[tbh]
	\begin{center}
		\includegraphics{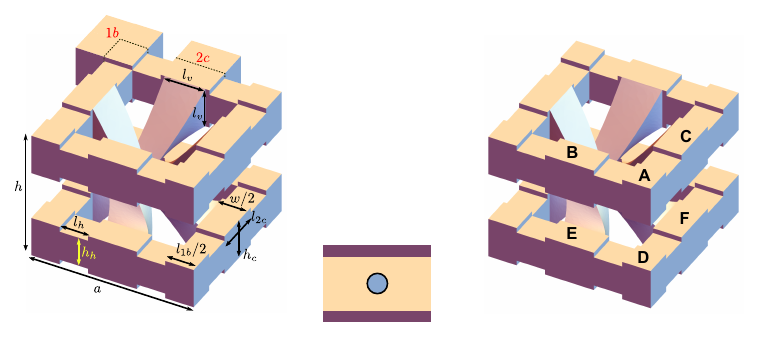}
	\end{center}
	\caption{{\bf Unit cell dimensions.} In the left panel, the major dimensions of the air-structure are indicated and detailed in Tab.~\ref{tab:SI-dimensions}. Note that the actual size of the cavities at $1a$ and $2c$ when arranged in a two dimensional lattice are shown for clarity. The unit cell boundaries are marked by the dashed line. The right panel shows the labelling of the cavities as used in the analysis below. The middle panel shows the relevant scale among the acoustic cavity (violet), the coupling channel (orange) and the microphone inserted into the sample (blue).}
	\label{fig:SI-UnitCell}
\end{figure}

The details of the unit cell's geometry of our acoustic crystal are summarized in Fig.~\ref{fig:SI-UnitCell} and Tab.~\ref{tab:SI-dimensions}. Note, that we provide as supplementary materials all relevant STL files from which the complete characterization of our sample can be extracted (\italicize{45}). 

\begin{table}[tbh]
  \begin{tabular}{llr}
    \hline
    \hline
  Variable & Explanation & Dimension in [mm]\\
  \hline
  $a$ & lattice constant & 40.0\\
  $h$ & total height & 32.0 \\
  $h_c$ & height of both cavities & 10.0\\
  $l_{1b}$ &length of cavity at $1b$ & 14.0\\
  $l_{2c}$ &length of cavity at $2c$ & 12.0\\
  $w$ & width of both cavities & 14.0\\
  $h_h$ & height of hopping channel & 7.5 \\
  $l_h$ & length of hopping channel & 7.0 \\
  $l_v$ & length/width of beginning of chiral hopping channel & 10.0\\
  \hline
  \hline
  \end{tabular}
  \caption{{\bf Unit cell dimensions.} Details to Fig.~\ref{fig:SI-UnitCell}.}
  \label{tab:SI-dimensions}
\end{table}

The sample shown in Fig.~1.G of the main text is printed on a Stratasys Connex Objet500 with PolyJet technology by Stratasys. The $300\,\mu{\rm m}$ resolution in the $xy$--plane and a resolution of $30\,\mu{\rm m}$ in the $z$--direction assure a fine surface finish for good hard-wall boundary conditions for the acoustic field, at least in the frequency/wavelength regime we are interested in: $\omega\approx 4{\rm kHz}$ and $\lambda\approx 8.6 {\rm cm}$).

The full sample consists of $L_x\times L_y=14\times14$ unit cells. It is divided into three main parts along the $x$ direction. The left and right parts (``bulk'' parts in the following) are composed of $6\times14$ unit cells. The central part (``cut'' part in the following and pink region of Fig.1.G of the main text) consists of $2\times14$ unit cells. Ten versions of the cut have been printed to perform twisted boundary conditions. Each part was constructed in three layers and later stacked and assembled. The printed material is VeroClear for the upper and lower layers. The middle layer is printed in VeroWhitePlus for the bulk parts. The materials used for the middle layers of each cut part are presented in Tab.~\ref{tab:SI-cutSpecs}.
\begin{table}[tbh]
  \begin{tabular}{rl}
    \hline
    \hline
  \multicolumn{1}{c}{$\lambda$}& Material \\
  \hline
  100\% & VeroWhitePlus\\
  67.5\% & VeroWhitePlus\\
  60.5\% & VeroWhitePlus\\
  46.5\% & VeroYellow \\
  39.5\% & VeroCyan \\
  32.5\% & VeroMagenta \\
  25.5\% & VeroGray \\
  18.5\% & VeroWhitePlus\\
  11.5\% & VeroWhitePlus\\
  0.0\% & VeroBlackPlus \\
  \hline
  \hline
  \end{tabular}
  \caption{{\bf Details on cuts for twisted boundary conditions.} $\lambda$ is the degree of separation, where $\lambda=1$ is the unperturbed sample and $\lambda=0$ corresponds to two disconnected halves.}
  \label{tab:SI-cutSpecs}
\end{table}

\subsection{Measurement and signal analysis.} 
\label{section:signal}
The acoustic signals are generated with speakers SR-32453-000 from Knowles. The pressure fields are measured via a sub-wavelength microphone FC-23629-P16 from Knowles with a diameter of $2.6\,{\rm mm}$ which is mounted on a $2\,{\rm mm}$ steel rod to scan the inside of the acoustic crystal. The rod is inserted into the sample with the help of a stage LRQ600 from Zaber to acquire a full bulk spectroscopy. The diameter of the rod is sufficiently small compared to the channels width and the acoustic wavelength to ensure a negligible perturbation of the signal.  

\begin{figure}[tbh]
	\begin{center}
		\includegraphics{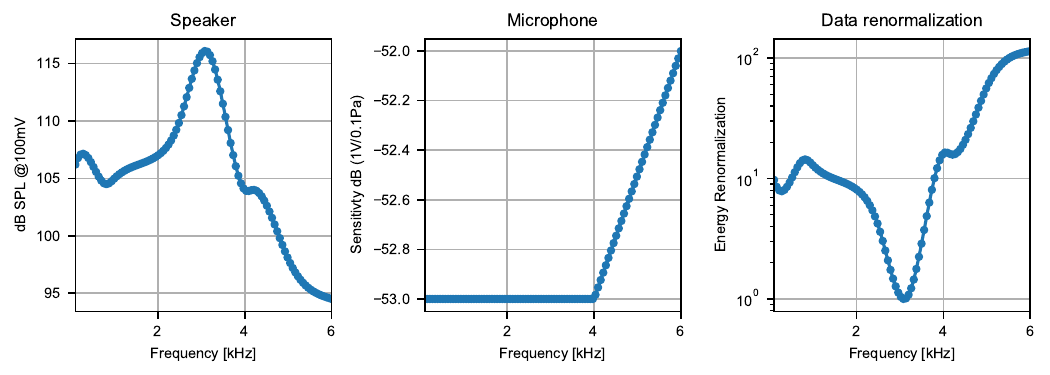}
	\end{center}
	\caption{{\bf Speaker and microphone responses.} In the left panel, frequency response as sound pressure level in ${\rm dB}$ of the speaker SR-32453-000 from Knowles using $100{\rm mV}$ drive signal in the frequency range $0$--$6 {\rm kHz}$. The middle panel shows the sensitivity in ${\rm dB}$ of the microphone FC-23629-P16 from Knowles relative to $1 {\rm V}/0.1 {\rm Pa}$. In the right panel, renormalization factor applied to the data as a function of frequency obtained from the convolution of the microphone and speaker responses.}
	\label{fig:SI-Renormalization}
\end{figure}

For the bulk spectrum, we measure 394 frequency points between $0.1$ and $6\,{\rm kHz}$ to obtain phase and amplitude information using a lock-in amplifier. The amplitudes have been normalized to account for the response of both the speaker and the microphone at different frequencies, cf. Fig.~\ref{fig:SI-Renormalization}. In unit cell coordinates $(i_x,i_y)$ running from $i_x,i_y=0,\dots,13$, we keep the exciter fixed at the $D$--site of unit cell $(6,7)$, cf. Fig.~\ref{fig:SI-UnitCell}. The crystal is then scanned at the center of all the sub-lattice sites $A,B,C,D,E,F$. The data displayed in Fig.~2.A of the main text are based on bulk measurement. The first two measurement points closer to any edges have not been taken into consideration to avoid spurious surface effects. 

For the analysis of modes localized at the symmetry center of the cut, we measure 200 frequency points between $3.2$ and $4.6\,{\rm kHz}$, i.e., around the gap frequency range. We excite at $(5,7)$ on the $D$ site and we measure the local density of state at the same location of the speaker. The dataset for each cut has been individually normalized with respect to its maximum value. 

Finally, for the study of the corner modes of Sec.~\ref{section:corner}, we use the same frequency range but excite from $(4,0)$ and $(9,0)$ on site A. For this set of measurements, only the $A$ and $C$ sub-lattices have been scanned over a $2\times 2$ unit cells grid on both sides of the full cut that disconnects the sample into two halves ($\lambda=0.0\%$), see Fig.~\ref{fig:SI-CornerMode}.A. 

\section{Supplementary Text}
\subsection{Fragile topology and topological quantum chemistry} \label{section:tqc}
Crystalline structures with translational symmetry are usually described in terms of Bloch wavefunctions (eigenvectors) and their band structure (eigenvalues) in momentum space. The wavefunctions form irreducible representations of the little group at momentum ${\bf k}$, defined as:
\begin{linenomath}
  \begin{equation}
     G_{\bf k}=\lbrace g=\{p_g|{\bf r}_g\}\in G \,|\, p_g{\bf k}\approx{\bf k} \rbrace,
  \end{equation}
\end{linenomath}
where $G$ is the crystal space group, $p_g$ is the point group part of $G$, ${\bf r}_g$ is the translation part of $G$, and $\approx$ indicates equivalence modulo translation by integer multiples of reciprocal lattice vectors. Note that $G_{\bf k}$ contains all the translation symmetries.
Topological quantum chemistry (TQC) (\italicize{10}, \italicize{24}) maps this momentum space description to a real space picture and offers simple criteria to asses the topology of a set of bands.

A Wyckoff position is a generic location inside the unit cell of a crystal. Each Wyckoff position $w$ has its own site-symmetry group $G_w$ which is a subgroup of the crystal space group $G$ and is defined as:
\begin{linenomath}
  \begin{equation}
     G_w=\lbrace g=\{p_g|{\bf r}_g\}\in G \,|\, p_g{\bf r}_w+{\bf r}_g={\bf r}_w \rbrace,
  \end{equation}
\end{linenomath}
where ${\bf r}_w$ is the location of the Wyckoff position. Among all points inside the unit cell, maximal Wyckoff positions play an important role. These are defined as sites where the site-symmetry group is a maximal subgroup of $G$, i.e.,
\begin{linenomath}
  \begin{equation}
     \nexists \,H \,|\, G_w \subset H \subset G. 
  \end{equation}
\end{linenomath}
Band representations (BRs) (\italicize{25}) are particular group representations that are induced by the irreducible representation of the site-symmetry group at the Wyckoff positions. 
Elementary band representations (EBRs) are BRs that cannot be further decomposed into multiples of BRs. These are induced by localized orbitals at maximal Wyckoff positions. \\

\fbox{
 \parbox{0.9\textwidth}{%
{\bf A central statement of TQC is that whenever a set of bands has symmetry properties that are not compatible with the ones of a sum of EBRs, these bands are topological. In fact, it means they cannot be adiabatically connected to an atomic limit with localized orbitals at lattice sites.} 
}
}
\\
\vspace{5pt} 

The irreducible representations at different high symmetry points (HSPs) are not independent, but related by certain compatibility relations, cf. Table~\ref{tab:compRelD} for the wallpaper group $p4mm$ (\italicize{23}, \italicize{46}).\\

\fbox{
 \parbox{0.9\textwidth}{%
{\bf From a crystalline perspective (\italicize{47}), the topology of a group of bands is entirely characterized by the symmetry data vector $B$ containing the multiplicity of the various irreducible representations at different HSPs.} 
}
}
\\
\vspace{5pt}

Any band structure with symmetry data $B$ that satisfies the compatibility relations can always be written as a linear combination of EBRs:
\begin{linenomath}
  \begin{equation}
    \label{eqn:decomposition}
    B=\sum_ip_i \,\text{EBR}_i,
  \end{equation}
\end{linenomath}
where $\text{EBR}_i$ is the symmetry data vector of the $i^\text{th}$ EBR of the group under consideration, see Table~\ref{tab:p4mmmEBR} for all the EBRs of $p4mm$. The decomposition of Eq.~\eqref{eqn:decomposition} is generically not unique for a given $B$. However, if one considers all possible decompositions, the coefficients $p_i$ can determine the topological nature of the bands. Three different scenarios are possible:
\begin{enumerate}
  \item $\exists\, p_i\, |\, p_i\notin \mathbb{Z} \quad \forall\; \text{decompositions}$,
  \item $\forall  p_i\,, p_i \in \mathbb{Z}\; \text{and}\; \exists p_i\notin \mathbb{N} \quad \forall\; \text{decompositions}$,
  \item $\forall  p_i\,, p_i \in \mathbb{N} \quad \forall\; \text{decompositions}$,
\end{enumerate} 
where $\mathbb{Z}$ ($\mathbb{N}$) indicates the set of integer (natural) numbers. Case 3 corresponds to insulators compatible with trivial insulators, although one needs to check Berry phases for a conclusive statement on the trivial nature of these bands. The other cases signal topological insulators. In particular, case 1 indicates stable topological insulators, while case 2 fragile ones (\italicize{15}--\italicize{17}, \italicize{22}). 

In Sec.~\ref{section:tbModel} below, we will decompose the bands of our acoustic system according to the scheme outlined here.

\subsection{Wallpaper group $p4mm$}
\label{section:p4mm}
In this work, we focus our attention on the wallpaper group $p4mm$. A generic group element $g=\{p_g|{\bf r}_g\}$ contains a point group part $p_g$ and a translational part ${\bf r}_g$. Here we are interested in the former, which in the case of $p4mm$ is isomorphic to the point group $C_{4v}$ (alternatively $4mm$). The generators of $C_{4v}$ are the identity $\mathbb{I}$, a two-fold rotation $c_2$, a four-fold rotation $c_4$ and a mirror $m_{10}$. Here, $m_{ij}$ is defined as the mirror with respect to the plane perpendicular to the vector $i{\bf a}_x+j{\bf a}_y$ and $\bar{\imath}=-i$. 
For the sake of clarity, we describe the actions of these symmetries on the basis vectors of a square lattice, ${\bf e}_1={\bf\hat x}$ and ${\bf e}_2={\bf\hat y}$:
\begin{align}
	\mathbb{I}:\quad& ({\bf e}_1,{\bf e}_2) \rightarrow (\phantom{-}{\bf e}_1,\phantom{-}{\bf e}_2) , \\
	c_2:\quad& ({\bf e}_1,{\bf e}_2) \rightarrow (-{\bf e}_1,-{\bf e}_2) ,\\
	c_4:\quad& ({\bf e}_1,{\bf e}_2) \rightarrow (\phantom{-}{\bf e}_2,-{\bf e}_1) ,\\
  m_{10}:\quad& ({\bf e}_1,{\bf e}_2) \rightarrow (-{\bf e}_1,\phantom{-}{\bf e}_2) . 
\end{align}
Further symmetries of $p4mm$ include the mirrors $m_{01}$, $m_{11}$, $m_{\bar{1}1}$ and additional glide reflections. A graphical description of these symmetries is shown in Figure~1.B of the main text. In Sec.~\ref{section:tbModel} we provide the matrix representation of these symmetries for a concrete tight binding model.

The maximal Wyckoff positions of the group $p4mm$ are also labeled in Figure~1.B of the main text (\italicize{42}). The site $(0,0)$ is the symmetry center of the group and has a site-symmetry group isomorphic to $C_{4v}$. It is labeled as $1a$. An analogous situation is realized at the other four-fold rotation center $({\bf e}_1/2,\,{\bf e}_2/2)$: the $1b$ maximal Wyckoff position. Finally, site $({\bf e}_1/2,0)$ has site-symmetry group $C_{2v}$. The $c_4$ operator maps this point to $(0,{\bf e}_2/2)$ and they are both labeled as $2c$. This exhausts the maximal Wyckoff positions of the wallpaper group $p4mm$. A similar situation is realized in reciprocal space. Here, $M=(\pi,\pi)$ and $\Gamma=(0,0)$ have little group isomorphic to $C_{4v}$, while $X=(\pi,0)\,,(0,\pi)$ to $C_{2v}$. 

Real space orbitals transform according to one of the irreducible representations of the site-symmetry group of the position at which they are localized. In an analogous way, Bloch wavefunctions at momentum {\bf k} form irreducible representations of the little group at ${\bf k}$. We provide all the irreducible representations for the point group of $p4mm$ and its relevant sub-groups in Table~\ref{table:p4}. 
\begin{table}[b!th]
	\subfloat[$C_{4v}$ ($4mm$)]{
	\begin{tabular}[b]{cccccccc}
	    \hline
	    \hline   
	     & & &$\phantom{-}\mathds{1}$ & $\phantom{-}c_2$ & $\phantom{-}m_{10}$ & $\phantom{-}m_{11}$ & $\phantom{-}c_4$ \\
	    \hline
	    $A_1$&$K_1$ &\includegraphics[height=8pt]{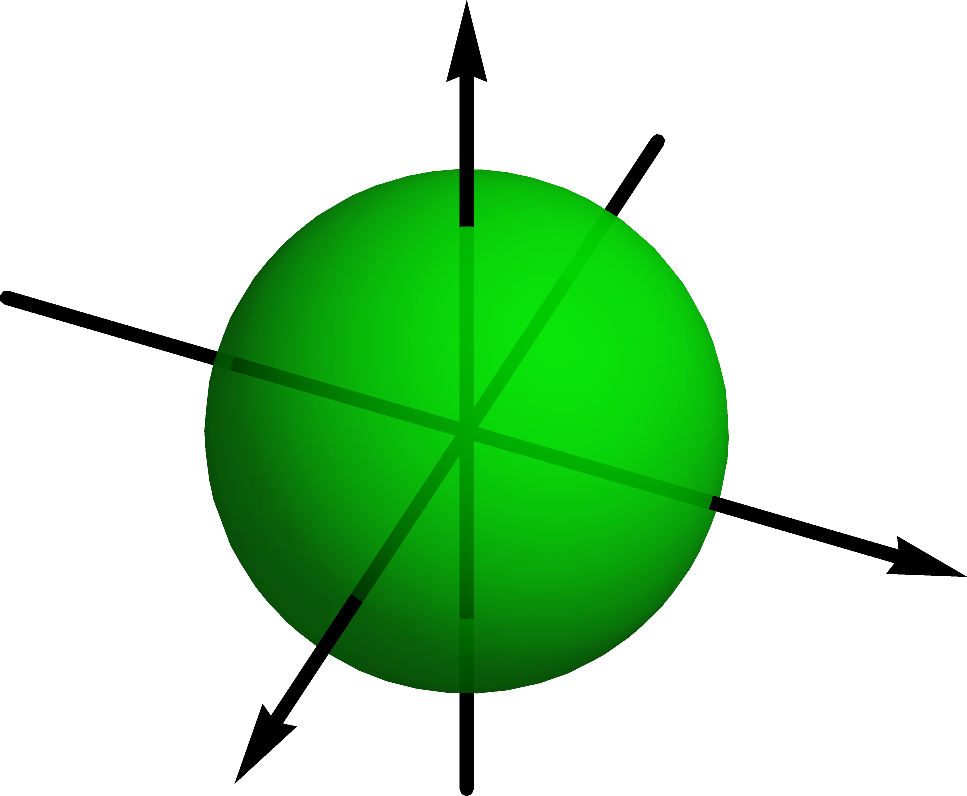}  & $\phantom{-}1$ & $\phantom{-}1$ & $\phantom{-}1$ & $\phantom{-}1$ & $\phantom{-}1$\\
	    $B_1$&$K_2$& \includegraphics[height=8pt]{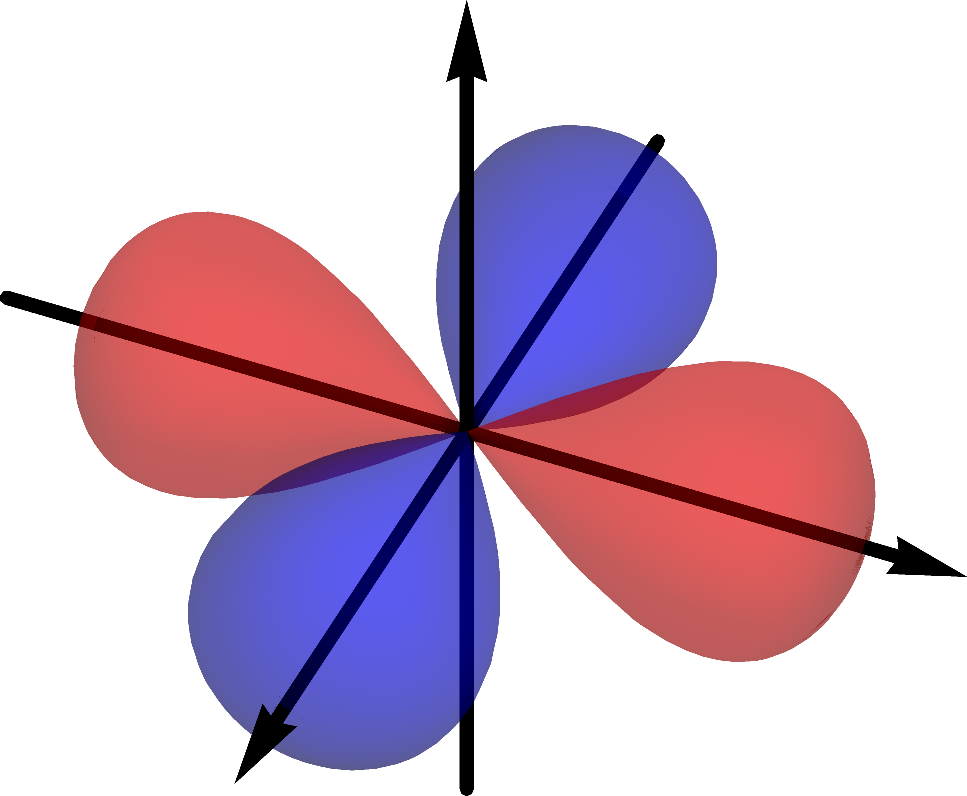} & $\phantom{-}1$ & $\phantom{-}1$ & $\phantom{-}1$ & $-1$ & $-1$ \\
	    $B_2$&$K_3$& \includegraphics[height=8pt]{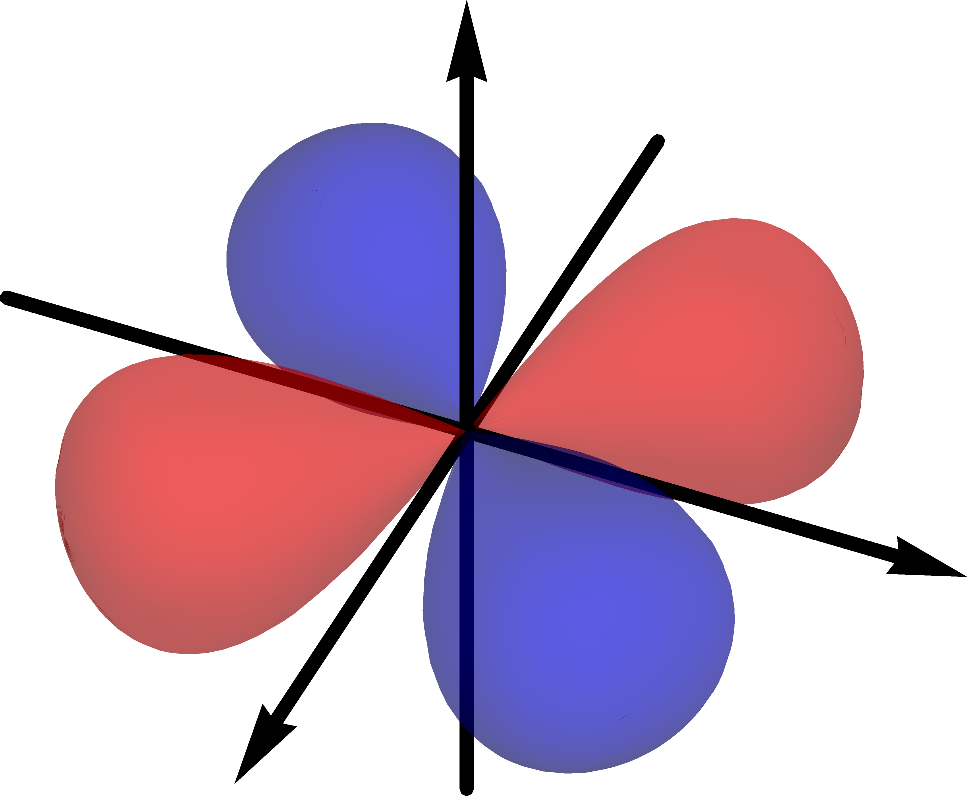} & $\phantom{-}1$ & $\phantom{-}1$ & $-1$ & $\phantom{-}1$ & $-1$ \\
	    $A_2$&$K_4$& \includegraphics[height=8pt]{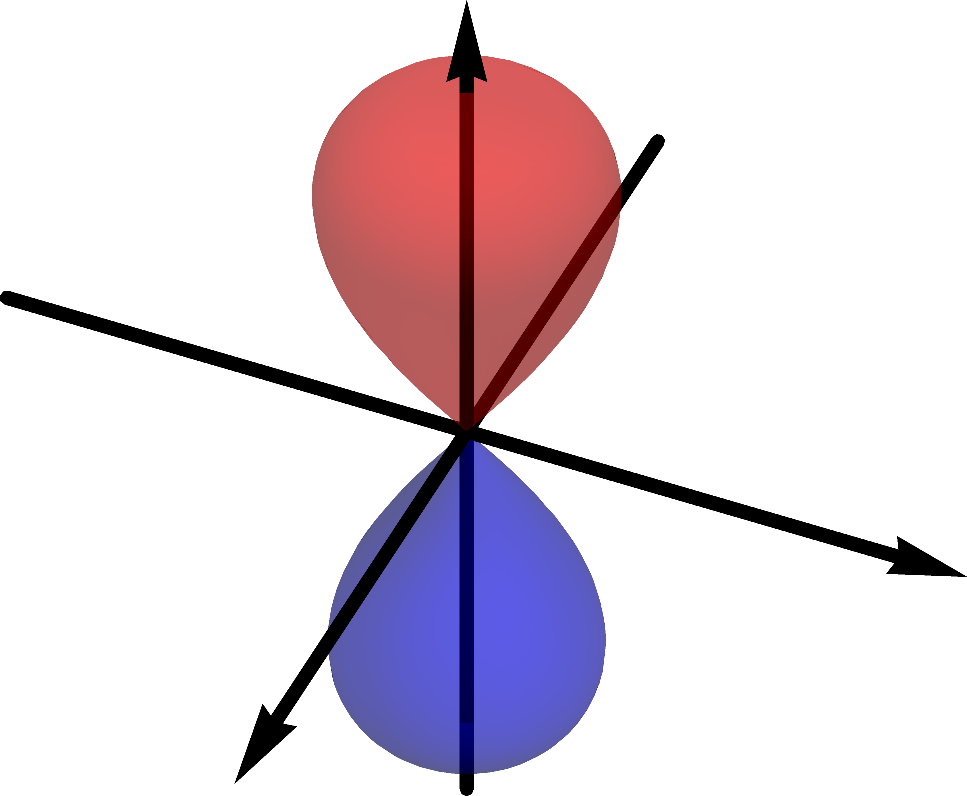} & $\phantom{-}1$ & $\phantom{-}1$ & $-1$ & $-1$ & $\phantom{-}1$\\
	    $E$&$K_5$& \includegraphics[height=8pt]{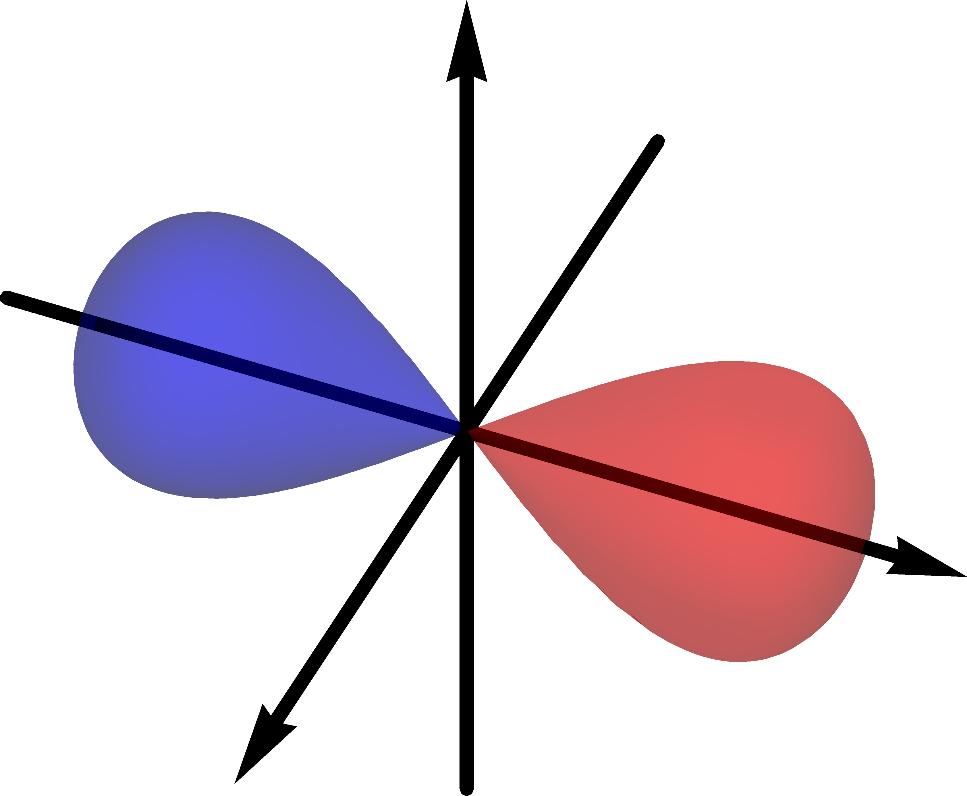}\includegraphics[height=8pt]{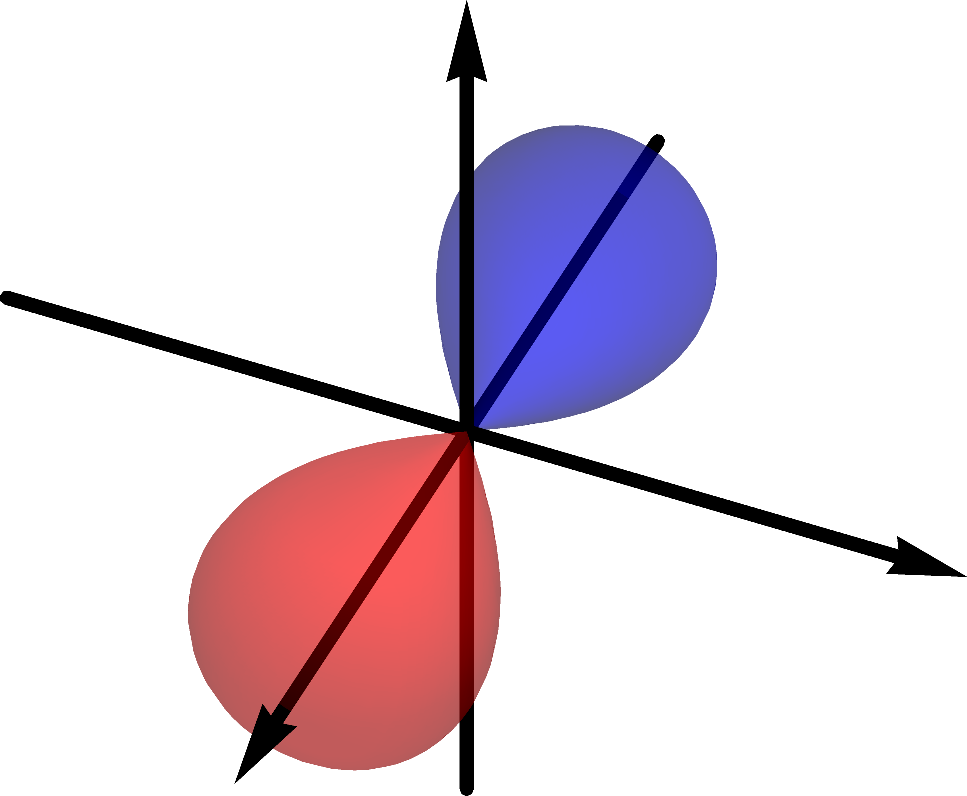} & $\phantom{-}2$ & $-2$ & $\phantom{-}0$ & $\phantom{-}0$ & $\phantom{-}0$ \\
	    \hline
	    \hline  
	\end{tabular}
	}
	\qquad
	\subfloat[$C_{2v}$ ($2mm$)]{%
	\begin{tabular}[b]{ccccccc}
	    \hline
	    \hline   
	     & & &$\phantom{-}\mathds{1}$ & $\phantom{-}c_2$ & $\phantom{-}m_{10}$ & $\phantom{-}m_{01}$ \\
	    \hline
	    $A_1$&$K_1$ &\includegraphics[height=8pt]{./A1}  & $\phantom{-}1$ & $\phantom{-}1$ & $\phantom{-}1$ & $\phantom{-}1$ \\
	    $A_2$&$K_2$& \includegraphics[height=8pt]{./A2} & $\phantom{-}1$ & $\phantom{-}1$ & $-1$ & $-1$ \\
	    $B_2$&$K_3$& \includegraphics[height=8pt]{./E1} & $\phantom{-}1$ & $-1$ & $\phantom{-}1$ & $-1$\\
	    $B_1$&$K_4$& \includegraphics[height=8pt]{./E2} & $\phantom{-}1$ & $-1$ & $-1$ & $\phantom{-}1$ \\
	    \hline
	    \hline  
	\end{tabular}
	}
	\qquad
	\subfloat[$C_{2}$ ($2$)]{%
	\begin{tabular}[b]{ccccc}
	    \hline
	    \hline   
	     & & &$\phantom{-}\mathds{1}$ & $\phantom{-}c_2$ \\
	    \hline
	    $A$&$K_1$ &\includegraphics[height=8pt]{./A1}  & $\phantom{-}1$ & $\phantom{-}1$ \\
	    $B$&$K_2$& \includegraphics[height=8pt]{./E1} & $\phantom{-}1$ & $-1$ \\
	    \hline
	    \hline  
	\end{tabular}
	}
    \caption{{\bf Irreducible representations of $\mathbf{p4mm}$.} Character tables and example orbitals for the point group of $p4mm$ and its relevant subgroups. The first column gives the standard name according to (\italicize{44}), the second column the label we give at the high-symmetry points in the Brillouin zone (e.g., $K_1\rightarrow \Gamma_1$ at the $\Gamma$--point), the third column depicts an example orbital in the respective irreducible representation. Note that, due to the non-trivial action of the mirror symmetries on the layer degree of freedom (see Sec.~\ref{section:tbModel}), a $p_z$ orbital transforms according to $A_2$ rather than $A_1$ in our system. For this reason, we refer to it as axial $p_z$ orbital.}
    \label{table:p4}
\end{table} 

The irreducible representations realized at different high symmetry points in momentum space are not independent. They are related by certain compatibility relations that ensure that different HSPs can be connected along high symmetry lines in momentum space with a spectral gap. For the wallpaper group $p4mm$, one considers: $\Delta=(k,0)$, $\Sigma=(k,k)$ and $Z=(\pi,k)$ for $k \in [0,\pi]$. Along these lines, the symmetry is reduced and the little groups are $m_{01}$, $m_{\bar{1}1}$ and $m_{10}$, respectively. The irreducible representations of these point groups are either even $K_1$ or odd $K_2$ under the mirror, where $K=\{\Delta,\Sigma,Z\}$. In Table~\ref{tab:compRelD}, we describe how the representations at HSPs get reduced along the high symmetry lines for the wallpaper group $p4mm$.

\begin{table}[!ht]
  \begin{tabular}{c|c|c|c|c|c}
    \hline
    \hline
  $\Gamma \rightarrow  \Delta$ & $X \rightarrow \Delta$ & $\Gamma
  \rightarrow \Sigma$ & $M \rightarrow \Sigma$ & $X \rightarrow Z$  & $M \rightarrow Z$  \\
  \hline
  $\Gamma_1 \rightarrow \Delta_1$ & $X_1 \rightarrow \Delta_1$ & $\Gamma_1 \rightarrow \Sigma_1$ & $M_1 \rightarrow \Sigma_1$ &  $X_1 \rightarrow Z_1$ & $M_1 \rightarrow Z_1$  \\
  $\Gamma_2 \rightarrow \Delta_1$ & $X_2 \rightarrow \Delta_2$ & $\Gamma_2 \rightarrow \Sigma_2$ & $M_2 \rightarrow \Sigma_2$ & $X_2 \rightarrow Z_2$ & $M_2 \rightarrow Z_1$  \\
  $\Gamma_3 \rightarrow \Delta_2$ & $X_3 \rightarrow \Delta_2$ & $\Gamma_3 \rightarrow \Sigma_1$ & $M_3 \rightarrow \Sigma_1$ & $X_3 \rightarrow Z_1$  & $M_3 \rightarrow Z_2$\\
  $\Gamma_4 \rightarrow \Delta_2$ & $X_4 \rightarrow \Delta_1$ & $\Gamma_4 \rightarrow \Sigma_2$ & $M_4 \rightarrow \Sigma_2$ & $X_4 \rightarrow Z_2$ & $M_4 \rightarrow Z_2$ \\
  $\Gamma_5 \rightarrow \Delta_1\oplus\Delta_2$ &    &$\Gamma_5  \rightarrow \Sigma_1\oplus\Sigma_2$  & $M_5 \rightarrow \Sigma_1\oplus\Sigma_2$ & & $M_5 \rightarrow Z_1\oplus Z_2$\\
  \hline
  \hline
  \end{tabular}
  \caption{{\bf Compatibility relations for $\mathbf{p4mm}$.} (\italicize{42}, \italicize{48}) The upper row indicates which HSP and high symmetry line are considered in each column. Each row presents how a particular irreducible representation at HSPs gets reduced along the high symmetry line.}
  \label{tab:compRelD}
  \end{table}
  
With the tool of topological quantum chemistry, we can relate the irreducible representations at HSPs to localized orbitals in real space that induce an elementary band representation. A complete list of the EBRs for $p4mm$ with time reversal symmetry and without spin-orbit coupling is given in Table~\ref{tab:p4mmmEBR} (\italicize{42}). 
\begin{table}[!ht]
  \begin{tabular}{lcccccccccccccc}
    \hline
    \hline
  & $(A_1)_{1a}$ & $(A_2)_{1a}$ & $(B_1)_{1a}$ & $(B_2)_{1a}$ & $(E)_{1a}$
  & $(A_1)_{1b}$ & $(A_2)_{1b}$ & $(B_1)_{1b}$ & $(B_2)_{1b}$ & $(E)_{1b}$
  & $(A_1)_{2c}$ & $(A_2)_{2c}$ & $(B_1)_{2c}$ & $(B_2)_{2c}$\\
  \hline
  $\Gamma$ & $\Gamma_1$ & $\Gamma_4$ & $\Gamma_3$ & $\Gamma_3$ & $\Gamma_5$ & $\Gamma_1$ & 
  $\Gamma_4$ & $\Gamma_2$ & $\Gamma_3$ & $\Gamma_5$ & $\Gamma_1\oplus\Gamma_2$ & $\Gamma_3\oplus\Gamma_4$ & $\Gamma_5$  & $\Gamma_5$ 
  \\
  $X$ & $X_1$ & $X_2$ & $X_1$ & $X_2$ & $X_3\oplus X_4$ & $X_3$ & 
  $X_4$ & $X_3$ & $X_4$ & $X_1\oplus X_2$ & $X_1\oplus X_3$ & $X_2\oplus X_4$ & $X_1\oplus X_4$  & $X_2\oplus X_3$ 
  \\
  $M$ & $M_1$ & $M_4$ & $M_2$ & $M_3$ & $M_5$ & $M_3$ & 
  $M_2$ & $M_4$ & $M_1$ & $M_5$ & $M_5$ & $M_5$ & $M_1\oplus M_2$  & $M_3\oplus M_4$ 
  \\
  \hline
  \hline
  \end{tabular}
  \caption{{\bf Elementary band representations for $\mathbf{p4mm}$.} (\italicize{42}) The upper row indicates the name of the EBR as $(K)_\ell$, where $K$ is the irreducible representation of the orbital that induces the EBR and $\ell$ its maximal Wyckoff position. Each column contains the irreducible representations at high symmetry points in momentum space for the different EBRs.}
  \label{tab:p4mmmEBR}
  \end{table}

\subsection{Twisted boundary conditions and spectral flow}
\label{section:twisted}
Twisted boundary conditions have been successfully applied to prove the quantization of the Hall conductance and the bulk-boundary correspondence of topological states (\italicize{49}, \italicize{50}). Here a generalized twisted boundary condition is used to detect a spectral signature of fragile topological insulators.

We consider a $C_2$--symmetric cut that divides our system in two regions: \RN{1} and \RN{2}. The twisted Hamiltonian $H(\lambda)$ is defined as:
\begin{linenomath}
\begin{equation}
\bra{\alpha} H(\lambda) \ket{\beta} = \begin{cases}
\phantom{\lambda}\bra{\alpha} H_0 \ket{\beta},\qquad & \alpha\in \RN{1}, \beta\in \RN{1}\\
\lambda\bra{\alpha} H_0 \ket{\beta},\qquad & \alpha\in \RN{1}, \beta\in \RN{2}\\
\lambda\bra{\alpha} H_0 \ket{\beta},\qquad & \alpha\in \RN{2}, \beta\in \RN{1}\\
\phantom{\lambda}\bra{\alpha} H_0 \ket{\beta},\qquad & \alpha\in \RN{2}, \beta\in \RN{2}
\end{cases},
\end{equation}
\end{linenomath}
where $H_0=H(1)$ is the original unperturbed Hamiltonian and $\lambda \in [-1,1]$. This amounts to multiplying by $\lambda$ all the hopping terms crossing the cut between region \RN{1} and \RN{2}. Note that during the twisting procedure, $C_2$ symmetry is preserved. By tuning $\lambda$ from $1$ to $-1$, one finds that $H(-1)=VH(1)V^\dagger$, where the gauge transformation $V$ is defined as:
\begin{linenomath}
\begin{equation}
  \label{eqn:gaugeCut}
V \ket{\alpha} = \begin{cases}
\phantom{-}\ket{\alpha} ,\qquad & \alpha\in \RN{1}\\
- \ket{\alpha},\qquad & \alpha\in \RN{2}
\end{cases}.  
\end{equation}
\end{linenomath}
$V$ anticommutes with $c_2$, i.e., $\left\{ V,c_2\right\}=0$. Namely, consider $\ket{\alpha}\in\RN{1}$. One can act first with $V$ and then $c_2$:  $c_2V\ket{\alpha}=c_2\ket{\alpha}=\ket{\alpha}\in\RN{2}$. On the other hand, reversing the order of operations: $Vc_2\ket{\alpha}=V\ket{\alpha}=-\ket{\alpha}\in\RN{2}$. This implies that if $\ket{\psi}$ is an eigenvalue of $H(1)$ and odd (even) under $C_2$, i.e., $c_2\ket{\psi}=\mp\ket{\psi}$, $V\ket{\psi}$ is an eigenvalue of $H(-1)$ even (odd) under $C_2$. In other words, if $\ket{\psi}$ transforms under even (odd) irreducible representation of $C_2$, $V\ket{\psi}$ will transform according to the opposite irreducible representation.

Following our companion paper (\italicize{34}), we would like to define a topological index that captures this change of symmetry representations under twisted boundary conditions. Since even and odd irreducible representations exchange their character under the procedure, we consider $\delta=m(B)-m(A)$, which counts the difference between the multiplicity of odd, $m(B)$, and even, $m(A)$, irreducible representations in the system, cf. Tab.~\ref{table:p4}.(c). If $H(1)$ has $\delta\neq 0$, then $\delta\rightarrow -\delta$ under twisted boundary conditions. The only way this can happen is via $\delta$ states flowing from the conduction band to the valence band and vice versa. The crossing of these states is protected by their opposite eigenvalues of $C_2$ symmetry. This robust spectral flow under twisted boundary conditions constitutes the sought-after spectral signature of fragile topology.

To compute the index $\delta$, one could analyze the symmetry representations in real space formed by the eigenvectors with energy below the gap for a system with open boundary conditions. A direct computation from the momentum space symmetry data would be preferable as it does not involve the diagonalization of large matrices. It turns out that this is possible for sufficiently large systems that preserve translational symmetry in the bulk. Consider a system with point group $C_4$, time reversal symmetry and no spin-orbit coupling (extra mirrors are not relevant for the following discussion, which can be directly applied to our model in wallpaper group $p4mm$). We can think of constructing a finite sample by adding few orbitals at the time. This process needs to preserve the underlying crystalline symmetry. Therefore, each time 4 orbitals that transform into each other under the generators of $C_4$ are brought into the systems from far away. To reflect the increasing dimensions of the Hilbert space, the symmetry representation of these four orbitals has to be: $A\oplus B \oplus E$. One sees that adding orbitals away from the symmetry center does not give rise to an imbalance in the multiplicity of different irreducible representations, i.e., $\delta=0$. The only way to get $\delta\neq 0$ is having orbitals at the symmetry center of the system as they transform back to themselves and are not subject to the constraint of forming the balanced representation $A\oplus B \oplus E$. This proves how only the irreducible representations induced by orbitals at the symmetry center of the model give non-trivial contributions to the index $\delta$.

Applying the tool of topological quantum chemistry (\italicize{10}), one can directly compute $\delta$ from the symmetry data at the high symmetry points in momentum space. In fact, this determines the EBR decomposition of the bands of interest. Among all the EBRs, only the ones induced by orbitals at the maximal Wyckoff position that corresponds to the symmetry center of the system are considered to compute the real space index $\delta$.

Let us briefly argue why the further distinction introduced by mirror symmetries can be neglected and why we consider a $C_2$--symmetric cut in a $C_4$--symmetric system. 
A more detailed discussion can be found in (\italicize{34}).
In the presence of time reversal symmetry and absence of spin orbit coupling, the irreducible representations of $C_{4v}$ can be reduced to $C_4$:
\begin{linenomath}
\begin{equation}
  \label{eqn:reduce4}
A_1 \downarrow C_4 =A, \quad A_2 \downarrow C_4 =A, \quad B_1 \downarrow C_4 =B, \quad  B_2\downarrow C_4 =B, \quad  E\downarrow C_4 =E\;.
\end{equation}
\end{linenomath}
The reductions of \eqref{eqn:reduce4} proves that the distinction introduced by the extra mirrors can be neglected as $m'(A)=m(A_1)+m(A_2)$, $m'(B)=m(B_1)+m(B_2)$ and $m'(E)=m(E)$, where the primed quantities indicate the multiplicity in $C_4$ rather than $C_{4v}$. However, the cut breaks $C_4$ and preserves only $C_2$. It is then relevant to see how irreducible representations of $C_4$ reduce to $C_2$. $A$ and $B$ irreducible representations of $C_4$ have $C_2$ character $+1$. On the other hand, $E$ has character $-1$ under the same symmetry. Therefore we can provide a formula for the $\delta$ index in terms of $C_4$ irreducible representation:
\begin{linenomath}
\begin{equation}
  \label{eqn:index4}
\delta=m'(B)-m'(A)=2m(E)-m(A)-m(B)\;,
\end{equation}
\end{linenomath} 
where the primed multiplicities refer to $C_2$.

In the example of our model, the EBR decomposition of the two lower bands is (see next section):
\begin{linenomath}
\begin{equation}
  \label{eqn:EBRFlow}
(A_1)_{1b}\oplus (A_2)_{2c} \ominus (B_2)_{1a}\;.
\end{equation}
\end{linenomath}
The symmetry center is located at the $1a$ maximal Wyckoff position. The EBR induced by a localized orbital at $1a$ is even under inversion symmetry and therefore our system has $\delta=1$. Hence, we expect symmetry enforced flow of one even state from the upper bands to the bands below the gap under twisted boundary conditions. The extra minus sign in $\delta$ comes from the negative coefficient in the EBR decomposition.
This fact does not alter the conclusion. In fact, this decomposition is not unique and alternative ones where all the EBRs induced from $1a$ have positive coefficients are possible, e.g., $(A_1)_{1a}\oplus (E)_{1a} \oplus (A_2)_{2c} \ominus (B_2)_{1b} \ominus (E)_{1b}$. An important result is that, although the EBR decomposition is not unique for a set of BRs, the index $\delta$ is (\italicize{34}).

In order to observe the spectral flow under twisted boundary conditions, certain conditions need to be satisfied (\italicize{34}). The twisted boundary needs to be $C_2$ symmetric and hence pass through the symmetry center of the system. At the same time, to have the well defined gauge transformation of Eq.~\eqref{eqn:gaugeCut}, the cut cannot go through orbitals but only through the couplings. This, in turn, requires that the EBR-decomposition of the fragile bands contains an EBR induced by a localized orbital at the symmetry center. At the very same location, however, there cannot be any physical orbital of the system. This situation is realized in our model, where the symmetry center is located at $1a$, while all the lattice sites (or acoustic cavities for the experimental samples) are located at $1b$ and $2c$ sites.

The twisting procedure requires to adiabatically tune $\lambda$ from $1$ to $-1$. Couplings of opposite signs are not natural in low frequency acoustics and might suggest that the spectral signature described here is out of reach for sufficiently simple acoustic metamaterials. Nevertheless, there is an extra symmetry that assures how tuning $\lambda: 1\rightarrow 0$ suffices to detect spectral flow: the spectrum is symmetric under $\lambda\rightarrow -\lambda$ since $H(-\lambda)=VH(\lambda)V^\dagger, \quad \forall \lambda \in \left[0,1\right]$.

\subsection{Tight binding model and symmetry analysis}
\label{section:tbModel}
We introduce a tight binding model for a bilayer Lieb lattice with chiral couplings between the two inequivalent two-fold connected sites of different layers. The primitive lattice vectors are ${\bf a}_x=(a,0)$ and ${\bf a}_y=(0,a)$, where in the following we set $a=1$. The basis is organized as $\psi=(\psi_A,\psi_B,\psi_C,\psi_D,\psi_E,\psi_F)$, where the labels refer to the cavities of Fig.~\ref{fig:SI-UnitCell} ($A$ and $D$ at maximal Wyckoff position $1b$ and $B$,$C$,$E$ and $F$ at $2c$ ), and the Hamiltonian is:
\begin{linenomath}
\begin{equation}
  \label{eqn:tbModel}
  H(k_x,k_y)=\begin{psmallmatrix}
  m & t_0(1+e^{ik_x}) & t_0(1+e^{ik_y}) & 0 & 0 & 0\\
  t_0(1+e^{-ik_x}) & 0 & 0 & 0 & 0 & \chi(1+e^{-i(k_x-k_y)})\\
  t_0(1+e^{-ik_y}) & 0 & 0 & 0 & \chi(e^{ik_x}+e^{-ik_y})& 0 \\
  0 & 0 & 0 & m &t_0(1+e^{ik_x}) & t_0(1+e^{ik_y}) \\
  0 & 0 & \chi(e^{-ik_x}+e^{ik_y}) & t_0(1+e^{-ik_x}) & 0 & 0 \\
  0 & \chi(1+e^{i(k_x-k_y)}) & 0 & t_0(1+e^{-ik_y}) & 0 & 0 \\
  \end{psmallmatrix}\,.
\end{equation}
\end{linenomath}
The acoustic crystal is far from a discrete tight binding limit. Nevertheless, the simple model of Eq.~\eqref{eqn:tbModel} helps to gain further insights on the experimental system under study.

The model of Eq.~\eqref{eqn:tbModel} has group symmetry $p4mm$. The generators of the point group part have been introduced in Sec.~\ref{section:p4mm}. Here, we present the matrix representation of the relevant symmetries and their action in momentum space for the system under consideration. The symmetry center is taken to be the site $1a$ of the unit cell.

\begin{linenomath}
\begin{equation}
  \mathbb{I} = \mathbb{I}_{2\times 2}\otimes \mathbb{I}_{3\times 3}
 \qquad (k_x,k_y)\rightarrow (k_x,k_y)\,,
  \end{equation}
\end{linenomath}

\begin{linenomath}
\begin{equation}
  c_2 = \mathbb{I}_{2\times 2}\otimes\begin{pmatrix}
   e^{i(k_y+k_x)} & 0 & 0  \\
  0 & e^{ik_y} & 0  \\
  0 & 0 & e^{ik_x}\\
  \end{pmatrix} \qquad (k_x,k_y)\rightarrow (-k_x,-k_y)\,,
  \label{eqn:one}
  \end{equation}
\end{linenomath}
  
\begin{linenomath} 
\begin{equation}
 c_4=\mathbb{I}_{2\times 2}\otimes\begin{pmatrix}
  e^{ik_x} & 0 & 0  \\
  0 & 0 & e^{ik_x} \\
  0 & 1 & 0  \\
  \end{pmatrix} \qquad (k_x,k_y)\rightarrow (k_y,-k_x) \,,
  \label{eqn:two}
  \end{equation}
\end{linenomath}

\begin{linenomath}
\begin{equation}
  m_{01}=\sigma_x\otimes\begin{pmatrix}
  e^{ik_y} & 0 & 0 \\
  0 & e^{ik_y} & 0 \\
  0 & 0 & 1 \\
  \end{pmatrix} \qquad (k_x,k_y)\rightarrow (k_x,-k_y)\,,
  \label{eqn:three}
  \end{equation}
\end{linenomath}

\begin{linenomath}  
\begin{equation}
  m_{10}=\sigma_x\otimes\begin{pmatrix}
   e^{ik_x} & 0 & 0 \\
  0 & 1 & 0 \\
  0 & 0 & e^{ik_x} \\
  \end{pmatrix} \qquad (k_x,k_y)\rightarrow (-k_x,k_y)\,,
  \label{eqn:four}
  \end{equation}
\end{linenomath}

\begin{figure}
	\begin{center}
		\includegraphics{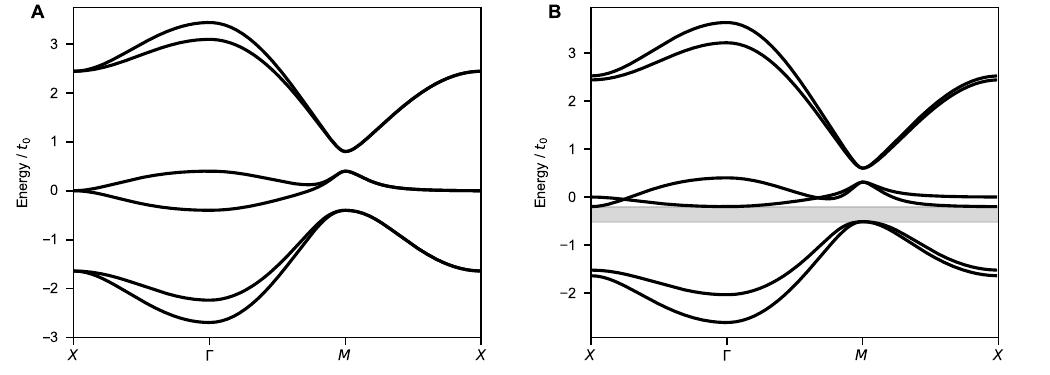}
	\end{center}
	\caption{{\bf Band structure of the tight binding model.} {\bf A} Spectrum of the tight binding model of Eq.~\eqref{eqn:tbModel} with $t_0=-1$, $m=0.8$ and $\chi=-0.2$. {\bf B} Spectrum for the same model with the addition of next-to-nearest neighbor couplings as in Eq.~\eqref{eqn:nnn} with $\mu=0.05$ and direct straight inter-layer $2c-2c$ coupling of strength $t'=-0.1$. The opened gap is shaded in grey.}
	\label{fig:SI-TB}
\end{figure}

\begin{linenomath}  
\begin{equation}
  m_{\bar{1}1}=\sigma_x\otimes\begin{pmatrix}
   1 & 0 & 0 \\
  0 & 0 & 1 \\
  0 & 1 & 0 \\
  \end{pmatrix}\qquad (k_x,k_y)\rightarrow (k_y,k_x)\,,
  \label{eqn:five}
  \end{equation}
\end{linenomath}
where $\sigma_i$ are Pauli matrices acting on the layer degree of freedom. Note that the mirror symmetries exchange the layer degree of freedom. This is why a $p_z$ orbital transforms according to the irreducible representation $A_2$ rather than $A_1$, see Table~\ref{table:p4}. To avoid confusion, we refer to it as axial $p_z$ orbital.  
The band structure of the model is shown in Fig.~\ref{fig:SI-TB}.A for $t_0=-1$, $m=0.8$, and $\chi=-0.2$. The six bands split into three branches. The gap of our interest is the one between the lower two branches. The observed double degeneracy along $M-X$ is enforced by an accidental symmetry at the boundary of the first Brillouin zone (BZ). In fact, one can define the operator 
\begin{linenomath}  
\begin{equation}
  \mathcal{O}=i\sigma_y\otimes\begin{pmatrix}
   1 & 0 & 0 \\
  0 & e^{-ik_x} & 0 \\
  0 & 0 & e^{-ik_y}\\
  \end{pmatrix}\mathcal{K}\,,
  \end{equation}
\end{linenomath}
where $\mathcal{K}$ indicates complex conjugation. Generically, $[H,\mathcal{O}]\neq 0$ and $\mathcal{O}$ is not a symmetry of the system. At the BZ boundaries, however, $[H,\mathcal{O}]= 0$. Given that $\mathcal{O}^2=-1$, the bands of this tight-binding model are all doubly degenerate at the BZ boundaries.

The irreducible representations realized by the model's bands at HSPs in the first BZ are presented in Table~\ref{tab:TBirreducible representations}.
\begin{table}[!ht]
  \begin{tabular}{lccc}
    \hline
    \hline
  & 1\&2 & 3\&4 & 5\&6 \\
  \hline
  $\Gamma$ & $\Gamma_1 \oplus \Gamma_4$ & $\Gamma_2  \oplus \Gamma_3$ & $\Gamma_1 \oplus \Gamma_4$ \\
  $X$ & $X_3  \oplus X_4$ & $X_1  \oplus X_2$ & $X_3  \oplus X_4$ \\
  $M$ & $M_5$ & $M_5$ & $M_2  \oplus M_3$\\
  \hline
  \hline
  \end{tabular}
  \caption{{\bf Irreducible representations at high symmetry points.} The first column indicates the high symmetry points considered. The second column contains the irreducible representations for bands 1\&2 of the model of Eq.~\eqref{eqn:tbModel} with $t_0=-1$, $m=1$ and $\chi=-0.2$. The third and fourth columns are for the irreducible representations of bands 3\&4 and 5\&6 of the same model, respectively.}
  \label{tab:TBirreducible representations}
  \end{table}

By inspecting the irreducible representations of the two lower bands of our system and the table of all EBRs of $p4mm$, cf. Table~\ref{tab:p4mmmEBR}, we confirm that they decompose as:
\begin{linenomath}
  \begin{equation}
    (A_1)_{1b}\oplus (A_2)_{2c} \ominus (B_2)_{1a}.
  \end{equation}
\end{linenomath}
A negative coefficient is present. One can further verify how a negative coefficient is always present for an arbitrary decomposition of these bands. According to the criteria of Sec.~\ref{section:tqc}, our bands realize fragile topology. We stress how this conclusion can be drawn entirely from the irreducible representations at high symmetry points in momentum space.

The addition of direct straight coupling between $2c$ sites of different layers removes the double degeneracy at the boundary of the BZ, cf. Fig.~\ref{fig:SI-TB}.B. This explains the lacks of degeneracy observed in the finite-element simulations and experimental measurements of the acoustic metamaterial. The degeneracy at the $M$ point is enforced by the two dimensional irreducible representation $M_5$ realized by the bands and is observed also in the experimental data. 

This tight binding model lacks of a direct gap in its spectrum. The energy of the third band at $\Gamma$ is equal to the energy of the first and second bands at $M$. For our purposes, it is preferable to have a full gap through the whole spectrum. This can be obtained adding next-to-nearest neighbor couplings to the Hamiltonian: $H'=H+H_{NNN}$, where
\begin{linenomath} 
  \begin{equation} 
   H_{NNN}=\mathbb{I}_{2\times 2}\otimes\begin{pmatrix}
  2t'\left(\cos{(k_x)}+\cos{(k_y)}\right) & 0 & 0 \\
  0 & 2t'\cos{(k_x)} & 0 \\
  0 & 0 & 2t'\cos{(k_y)} \\
  \end{pmatrix}\,. 
  \label{eqn:nnn}
  \end{equation}
\end{linenomath} 
The spectrum in the presence of next-to-nearest neighbor couplings better resembles the measured bands of the sonic crystal, cf. Fig.~\ref{fig:SI-TB}.B. This further confirms that the latter is far from a tight binding limit with only nearest neighbor hopping terms.

\subsection{Symmetry analysis from experimental data} 
\label{section:symmetry}
The symmetries of the bands can be extracted directly from the measurements. The results are presented in Fig.~2 of the main text. Here, we provide further details on how those results have been obtained. 

As described in Sec.~\ref{section:signal}, we slide a sub-wavelength microphone into the phononic crystal to measure the pressure field at the center of all cavities (all lattice sites of the bilayer Lieb lattice). For each cavity we chirp the frequency of the exciting speaker (which resides at a fixed location, cf. Sec.~\ref{section:signal}) from  $0.1\textup{--}6$~kHz and record the response. Locking the measured signal to the exciter, we can extract phase and amplitude information of the acoustic field $\psi_\alpha(i,\omega)$, where $\alpha$ denotes one of the six cavities $A,\dots,F$ as shown in Fig.~\ref{fig:SI-UnitCell}, $i$ is the unit-cell coordinate and $\omega$ the frequency. In other words, we extract the Greens function $G_{\alpha,\beta}^{i,j}(\omega) =\langle \psi^*_\alpha(i,\omega)\psi_\beta(j,\omega)\rangle$.

For further analysis we take a Fourier transform 
\begin{equation}
G_{\alpha}({\bf k},\omega) = \frac{1}{\sqrt{N}} \sum_{i} e^{i {\bf k}\cdot ({\bf r}_i-{\bf r}_j)} G_{\alpha,\beta}^{i,j}(\omega) ,
\end{equation}
where $N$ is the number of unit cells considered. Note that we suppressed the indices $j$ and $\beta$ on the left hand side as we do not change the exciter location. The spectrum shown in Fig.~2.A of the main text shows $\sum_\alpha |G_\alpha({\bf k},\omega)|^2$, i.e, we add the response at the different sub-lattice sites $A,\dots,F$ incoherently. 

\begin{figure}
 \begin{center}
 	\includegraphics{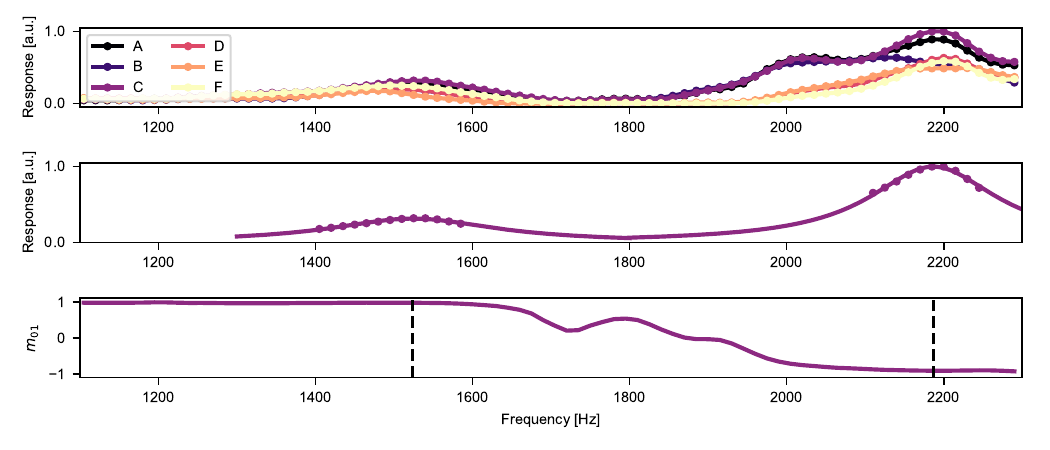}
 \end{center}
 \caption{{\bf Example of symmetry extraction.} In the top panel the response at ${\bf k}=(-0.2\pi,0)$ for each sub-lattice is plotted. In the middle panel, the fit of of the two highest peaks close to the expected modes frequencies is shown. In the bottom panel the expectation value $\langle\psi (\omega_n) |m_{01}| \psi(\omega_n)\rangle$ in the whole frequency range analyzed. The vertical dashed lines indicate the frequency of the modes from fit of the experimental data.}
 \label{fig:SI-Symm}
\end{figure}

The dispersion lines shown in Fig.~2.A of the main text and Fig.~\ref{fig:SI-comparison}.B are then extracted by fitting $|G_\alpha({\bf k},\omega)|^2$ for each ${\bf k}$ to a Lorentzian line-shape around the local maxima. This yields $\omega_n({\bf k})$ for the four lowest bands $n=1,\dots,4$. Note, that some modes might have zero or small weight on one of the six sublattices $\alpha=A,\dots, F$. We therefore fix $\omega_n({\bf k})$ via the fit to $|G_\alpha({\bf k},\omega)|^2$ with the largest weight to capitalize on the best signal-to-noise ratio in the determination of $\omega_n({\bf k})$. Fig.~\ref{fig:SI-Symm} shows an example for the two lower bands at ${\bf k}=(-0.2\pi,0)$. In this case, sub-lattice $\alpha=C$ shows the strongest response. 

Once $\omega_n({\bf k})$ is determined we can extract the symmetry eigenvalues. To this end, we write the Greens function as a vector

\begin{linenomath}
  \begin{equation}    \boldsymbol\psi({\bf k},\omega)=
	  \lbrace
	  G_A({\bf k},\omega),
	  G_B({\bf k},\omega),
	  G_C({\bf k},\omega),
	  G_D({\bf k},\omega),
	  G_E({\bf k},\omega),
	  G_F({\bf k},\omega)
	\rbrace\,.
  \end{equation}
\end{linenomath}
Note, that at $\omega\to\omega_n({\bf k})$ the vector $\boldsymbol\psi({\bf k},\omega_n({\bf k}))$ is directly proportional to the eigenmode of frequency $\omega_n({\bf k})$. We now compute the expectation values $\langle \boldsymbol\psi({\bf k},\omega) |\mathcal{A}|\boldsymbol\psi({\bf k},\omega)\rangle$, where $\mathcal{A}$ is one of the symmetry operators defined in Eqns.~(\ref{eqn:one}-\ref{eqn:five}). In the bottom panel of Fig.~\ref{fig:SI-Symm} we see how the expectation value changes from $+1$ to $-1$ when going from one band to the next (indicated by the dashed black lines).

\begin{figure}
 \begin{center}
 	\includegraphics{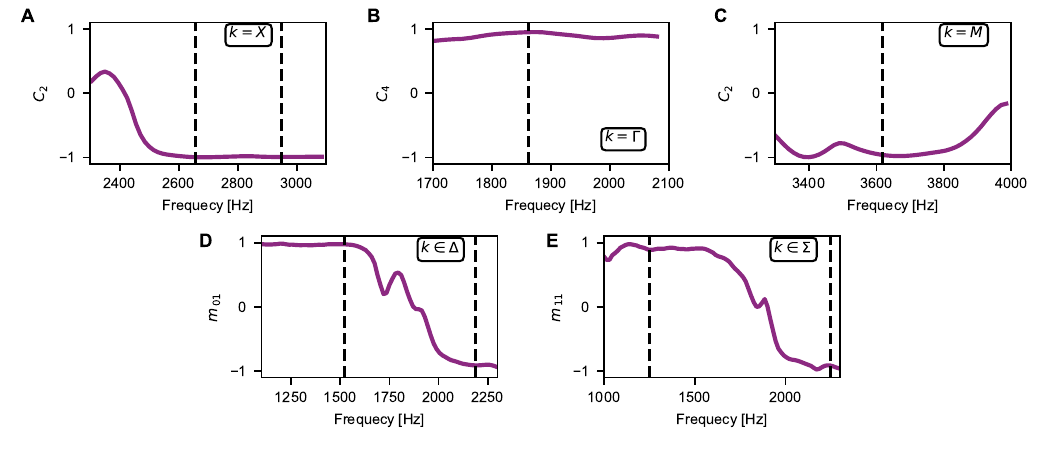}
 \end{center}
 \caption{{\bf Minimal set of complete symmetry extraction.} {\bf A} $C_2$ measurement at ${\bf k}=X$. {\bf B} $C_4$ measurement at ${\bf k}=\Gamma$. {\bf C} $C_{2}$ measurement at ${\bf k}=M$. {\bf D} $m_{01}$ measurement at ${\bf k}=(-0.2\pi,0)\in \Delta$. {\bf E} $m_{11}$ measurement at ${\bf k}=(-0.1\pi,-0.1\pi)\in \Sigma$. Black dashed lines indicate the frequency of the modes obtained by fit of the experimental data as described in Fig.~\ref{fig:SI-Symm} and Sec.~\ref{section:symmetry}.}
 \label{fig:SI-SymmMore}
\end{figure}

In order to fully determine the irreducible representations, we take advantage of the compatibility relations along high symmetry lines presented in Table~\ref{tab:compRelD} (\italicize{42}). In fact, along the lines $\Delta$ and $\Sigma$ we can choose momenta where the modes are largely separated in frequency and evaluate the mirror eigenvalues, cf. Fig.~\ref{fig:SI-SymmMore}.D and \ref{fig:SI-SymmMore}.E. This allows to infer the representations of the two lower bands by measuring $m_{10}$ and $m_{11}$ respectively: $\Delta_1\oplus\Delta_2$ and $\Sigma_1\oplus\Sigma_2$. A similar procedure cannot be carried out along $Z$ since the two modes are nearly degenerate. However, the mirror eigenvalues along the other high symmetry lines, together with the compatibility relations of Table~\ref{tab:compRelD}, uniquely determines the mirror eigenvalues at all high symmetry points. We then focus our attention on $c_2$ and $c_4$ operators at high symmetry points in momentum space. 

At $M$, we observe a stable degeneracy suggesting the realization of the two dimensional irreducible representation $M_5$. This observation is further substantiated by the negative eigenvalues of $c_2$, as can be seen in Fig.~\ref{fig:SI-SymmMore}.C. Table~\ref{table:p4} confirms that the only irreducible representation of $C_{4v}$, the little group at $M$, with negative eigenvalues of $c_2$ is $M_5$.

Even in the presence of nearly degenerate modes at $X$, $c_2$ has sharply quantized negative eigenvalues, as can be seen in Fig.~\ref{fig:SI-SymmMore}.A. This suggests that both modes have the same eigenvalue under $c_2$ and thus realize the representation $X_3\oplus X_4$ of the little group $C_{2v}$, cf. Table~\ref{table:p4}.

At $\Gamma$, the modes are spectrally separated and clearly non-degenerate. The measured positive eigenvalue of $c_4$ for the non-zero frequency eigenmode (see Fig.~\ref{fig:SI-SymmMore}.B), together with the compatibility relations $\Gamma\rightarrow\Delta$ and $\Gamma\rightarrow\Sigma$ of Table~\ref{tab:compRelD}, determines that the representation at $\Gamma$ is $\Gamma_1\oplus\Gamma_4$. 

Note that, with the compatibility relations of Table~\ref{tab:compRelD}, five symmetry measurements at five different momenta suffice to fully determine the representations realized at high symmetry points from experimental data. We call a set of symmetry operators and momentum points that allow to fully characterize the high symmetry points representations {\em complete}. The choice of this set is not unique and repeating the same analysis for a different one leads to an identical result.

An analogous analysis as outlined above for the lower two bands has been performed for the two bands above the gap. The results are in agreement with the expectations from the tight binding model and the finite-elements simulations, cf Table~\ref{tab:TBirreducible representations} and Fig.~\ref{fig:SI-comparison}.C. 

Comparing the measured representations and the EBRs of Table~\ref{tab:p4mmmEBR}, we can decompose the bands of the acoustic metamaterial as:
\begin{align}
	\text{bands 1\&2} &: (A_1)_{1b}\oplus (A_2)_{2c} \ominus (B_2)_{1a}, \\
	\text{bands 3\&4}  &: (B_2)_{1a}\oplus (A_1)_{2c} \ominus (A_1)_{1b} 
\end{align}

\subsection{Finite elements simulations of the acoustic crystal}

\begin{figure}
	\begin{center}
		\includegraphics{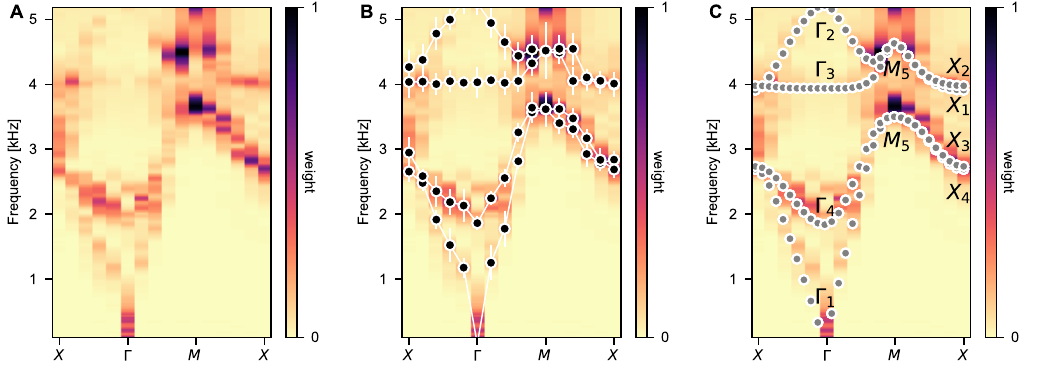}
	\end{center}
	\caption{{\bf Comparison between experiment and simulation.} {\bf A} Experimental results only. {\bf B} Fits to the local maxima overlaid on the experimental data. {\bf C} Overlay of the finite-elements simulations results. The labels of irreducible representation as obtained from numerical simulations are shown at high symmetry points.}
	\label{fig:SI-comparison}
\end{figure}

The linearized and dissipation-free approximation to the acoustic behavior in the sonic crystal can be efficiently simulated via finite-elements (FE). In Fig.~\ref{fig:SI-comparison}, we compare the measured dispersion, extracted as explained in Sec.~\ref{section:signal} and \ref{section:symmetry}, to the result of a finite-elements simulation performed in COMSOL Multiphysics. To allow for a bias free comparison, we show three panels: {\bf A} shows the experimental results only. Panel {\bf B} adds the fits to the local maxima that indicate the bands as measured in the experiment. Finally, in panel {\bf C}, we overlay the results of the finite-element simulation. 

Also the symmetry properties can be determined in the FE simulation. To this end, we consider a unit cell with periodic boundary conditions in all directions and sample the pressure field for high-symmetry momenta at 20 generic Wyckoff positions with orbit four inside the unit cell. We construct the symmetry operators as $20\times 20$ matrices inspecting how these 20 points transform into themselves under the symmetries action. The operators are then project onto the individual modes obtained from the FE simulation at fixed momentum. This procedure is repeated at all high symmetry momenta and for each of the four lowest modes. Consulting the character tables in Tab.~\ref{table:p4} for the little group of the respective high symmetry point ($C_{4v}$ at $\Gamma$ and $M$ and $C_{2v}$ at $X$), it is easy to determine the irreducible representations indicated in Fig.~\ref{fig:SI-comparison}.C and in Tab.~\ref{tab:TBirreducible representations}. They are in prefect agreement with the symmetry eigenvalues extracted from the measurements.

The same symmetry analysis can be performed in real space for a finite sample with open boundary conditions. Here, we sample the pressure field solution at the center of each acoustic cavity and the symmetry operators generically map cavities in one unit cell to a different unit cell. We evaluated the projection of the symmetry operators of $c_2$ and $c_4$ onto all states below the bulk gap frequency $3.7\,{\rm kHz}$. For a sample consisting of $5\times 5$ unit cells, this amounts to 95 states. A simple counting of eigenvalues leads to the result for the even, odd and doubly-degenerate irreducible representations of $C_4$ as
\begin{align}
	\#A &: 24,\\
	\#B &: 23,\\
	\#E &: 24,
\end{align}
as expected from the EBR decomposition, cf. Sec.~\ref{section:symmetry}. In fact, for the procedure of the twisted boundary conditions, we are only interested in the eigenvalues under two-fold rotations around the symmetry center of the cut, see Sec.~\ref{section:twisted}. Analysing the projected $c_2$ we find that we have one $c_2$--odd state more than we have $c_2$--even in the lowest two bands.

\subsection{Removal of non-topological surface modes}
\label{section:SCNstates}

Two dimensional fragile topological insulators can feature boundary states interpolating between conductance and valence bands. These states, however, lack the robustness of the analogous states in stable topological insulators. Namely, they can be gapped by edge perturbations that preserve the underlying crystalline symmetry.

The tight binding model of Eq.~\eqref{eqn:tbModel} has gapless edge modes crossing the bulk gap when placed on open boundary conditions. They are detrimental to our measurements, which require a full gap in the spectrum of the open system. A spin-Chern number can capture the presence of the boundary modes and suggest feasible routes to gap them (\italicize{51}, \italicize{52}).

To define a spin-Chern number for the model of Eq.~\eqref{eqn:tbModel}, one introduces a pseudospin $\tau_{\alpha}=\sigma_{\alpha}\otimes\mathbb{I}_{3\times 3}$, where the Pauli matrices $\sigma_{\alpha}$ act on the layer degree of freedom. None of the pseudospin components is conserved. One can still project $\tau_y$ (analogous results are obtained for the other components of the pseudospin degree of freedom) into the two bands below the lowest gap to split them. These split bands lead to well-defined fiber bundles which may carry a non-zero Chern number: the spin-Chern number. This is a well defined topological number (\italicize{51}). Nonetheless, it relies not only on the presence of a spectral gap, but also on a spin-projection gap. Moreover, it lacks of a well defined bulk-boundary correspondence: the bulk physics does not unambiguously determine what happens at the boundaries. At the interface between an insulator with non-zero spin-Chern number and vacuum, the spin-Chern number needs to vanish. This can occur in two distinct ways: the spectral gap closes at the boundaries or the spin-projection gap does. In the former case, gapless edge modes cross the energetic gap. In the latter, the spectral gap may remain open also at the boundaries (\italicize{53}).

\begin{figure}[tbh]
	\begin{center}
		\includegraphics{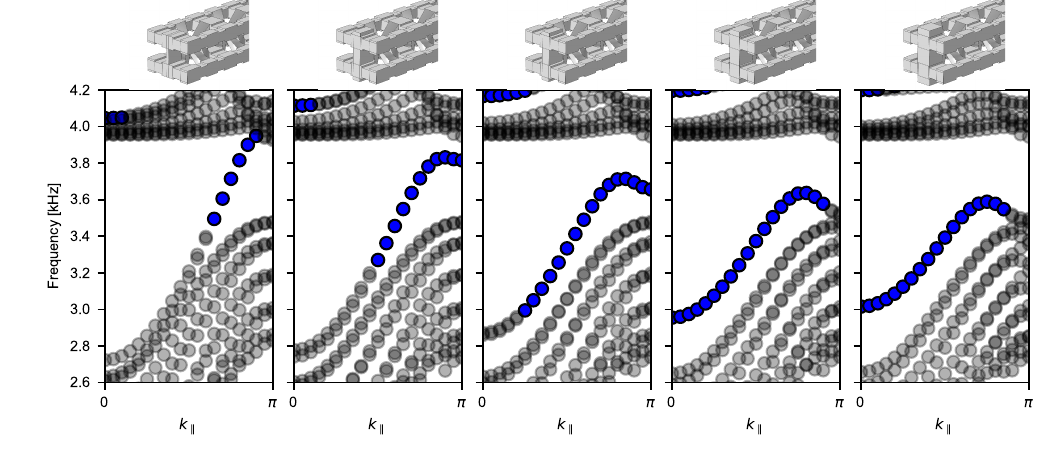}
	\end{center}
	\caption{{\bf Removal of non-topological surface modes.} From left to right the coupling between the $2c$--$2c$ cavities in the different layers is increased, which closes a spin-gap and leads to a removal of the surface modes. The color of the dots encodes whether the eigenstate is concentrated on the edge of a cylindrical geometry.
	}
	\label{fig:SI-SCNRemoval}
\end{figure}

To have a full gap in the spectrum of our sample, we add perturbations to the edges that close the spin-projection gap. A sufficiently strong direct coupling between the $2c-2c$ sites of different layers, i.e., from the $B$ cavity to the $E$ cavity and from $C$ to $F$, closes the spin-projection gap. In Fig.~\ref{fig:SI-SCNRemoval}, we substantiate via finite-elements simulations that an increasing coupling of this type opens a gap in the edge states of a cylindrical geometry.
As both finite-elements simulations and experimental measurements confirm, a full gap is induced in the system.

\subsection{Spectral flow: experimental characterization}
\label{section:flowMeasure}

As elaborated in Sec.~\ref{section:twisted}, fragile topology can induce a stable spectral signature in the form of spectral flow if the ``missing'' EBR is induced from a maximal Wyckoff position with no actual physical site. In this section, we analyze the spectral flow for a soft cut both in finite-elements simulations and in a tight-binding model.
\begin{figure}[tbh]
	\begin{center}
		\includegraphics{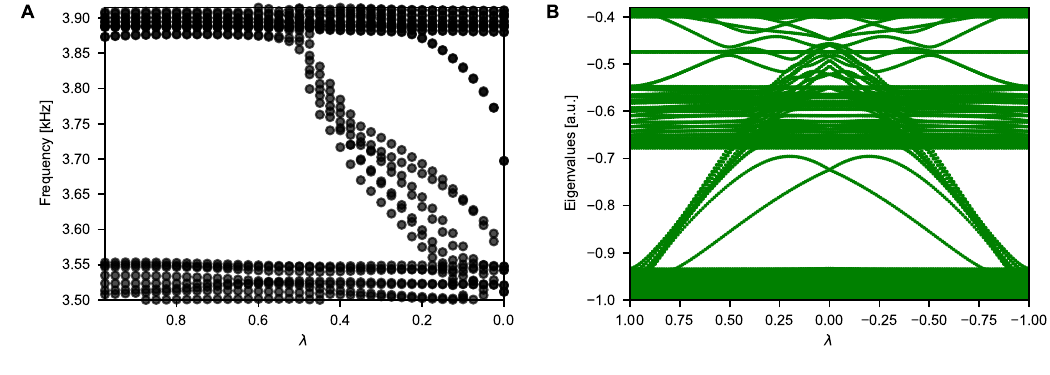}
	\end{center}
	\caption{{\bf Spectral flow for soft cut.} {\bf A} The finite-elements simulation of the acoustic crystal as a function of the cutting level $\lambda$. {\bf B} Spectral flow for a tight-binding model with the same topology. The parameters of the tight-binding model are $t_0=-1$, $\chi=-0.4$, $t'=0.2$, $m=2$, $q=-0.3$, and $\gamma=-0.6$, see text for details.}
	\label{fig:SI-FlowingSates}
\end{figure}

First, we induce the cutting procedure in the acoustic crystal. We slowly pinch off the hopping channels along the cut marked in Fig.~1.H of the main text. We do so by adding a rectangular obstruction of growing width. (See the attached STEP files and the inset of Fig.~1.G of the main text for the details of the geometry). 

In Fig.~\ref{fig:SI-FlowingSates} we present the numerical solution of the cutting procedure. Between $\lambda=0.5$ and  $\lambda=0$ a number of states that scales with the length of the cut is flowing through the bulk gap. The lowest of these flowing modes is centered at the cut and shown in Fig.~3.B of the main text. A corner mode is pulled into the gap at full separation $\lambda=0$, we further discuss this corner mode in the next section.

Let us first comment on the extensive nature of the flowing states. In Sec.~\ref{section:SCNstates}, we discuss how a pristine edge nucleates edge modes that connect the two bands. Moreover, we argued how one can remove them through the introduction of a direct $2c$--$2c$ layer-coupling. We remove these states also for the $C_2$--symmetric cutting procedure by adding direct $2c$--$2c$ hoppings between the layers along the cut. However, energetics are such that the so-removed spin-Chern number states spectrally overlap with the lower band. This can also be seen in Fig.~\ref{fig:SI-SCNRemoval}.

Hence, by our procedure we also move these states from the upper to the lower band. However, these are not protected by the fragility and/or $C_2$--symmetry. In particular they are ``$\delta$--neutral'': there are as many $C_2$--odd as $C_2$--even states flowing through the band-gap. It is the $C_2$--even state localized at the symmetry center of the cut which is protected by fragility.

We further verify this statement by calculating the symmetry eigenvalues  for all states below the gap at $\lambda=1$  and $\lambda=0$ and we indeed find that $\delta$ changes from $+1$ to $0$ during the cutting procedure.

As an additional verification, we show  in Fig.~\ref{fig:SI-FlowingSates}.B the numerical simulation of the soft cut for a tight-binding model, where we move the spin-Chern number band  on purpose to the  middle of the gap. We use the matrix
\begin{equation}
  H=\begin{psmallmatrix}
  m + t'(T_{x}+T_{y}+T_{-x}+T_{-y})  & t_0(1+T_{x}) & t_0(1+T_{y}) & 0 & 0 & 0\\
  t_0(1+T_{-x}) & q g_{xy} +t'(T_{x}+T_{-x})& 0 & 0 & \gamma f_{xy} & \chi(1+T_{-x+y})\\
  t_0(1+T_{-y}) & 0 & q g_{xy}+ t'(T_{y}+T_{-y}) & 0 & \chi(T_{x}+T_{-y})& \gamma f_{xy} \\
  0 & 0 & 0 & m +t'(T_{x}+T_{y}+T_{-x}+T_{-y})&t_0(1+T_{x}) & t_0(1+T_{y}) \\
  0 & \gamma f_{xy} & \chi(T_{-x}+T_{y}) & t_0(1+T_{-x}) & q g_{xy}+t'(T_{x}+T_{-x}) & 0 \\
  0 & \chi(1+T_{x-y}) & \gamma f_{xy} & t_0(1+T_{-y}) & 0 & q g_{xy}+t'(T_{y}+T_{-y})\\
  \end{psmallmatrix}\,.
\end{equation}
Here $T_{\pm x}$ ,$T_{\pm y}$ ,$T_{\pm x\pm y}$  denote the translation operators by one unit cell in the respective directions. The functions $f_{xy}$ and $g_{xy}$ select the sites [$f_{xy}$] and bonds [$g_{xy}$] along the cut. The parameters $\gamma$ and $q$ denote the $2c$--$2c$ interlayer hoppings and an on-site potential for the $2c$--sites, respectively. Finally, $t'$ is the next-nearest neighbor hopping amplitude, as introduced in~\eqref{eqn:nnn}. We detached the surface band from both the lower and upper bulk band. Regardless of all the many states leaving the lower bands, one can check one simple aspect: the index $\delta$ for states  below the eigenvalue $-0.9$ and $\lambda=1,0,-1$. One indeed finds that at $\lambda=0$,  $\delta=0$, whereas for $\lambda=\pm 1$ one obtains $\delta=\pm 1$ as expected. Moreover, one can check that the spin-Chern number states in the middle are $\delta$--neutral.

\subsection{Corner modes in fragile topological insulators} \label{section:corner}
Quantized fractional corner charges have been predicted as a possible spectral signature for fragile topological insulators in the electronic domain (\italicize{19}, \italicize{21}, \italicize{54}). In a charge neutral system, modes strongly localized at the corners are the counterpart of corner charges. Here, we address whether our system can be characterized in terms of its corner modes. 

The $C_4$--symmetry of the model constraints the value of the polarization (\italicize{54})
\begin{linenomath}
\begin{equation}
        \label{eqn:polarization}
      {\bf P}=p_x{\bf \hat{x}}+p_y{\bf \hat{y}}\,,
\end{equation}
\end{linenomath}
to $p_x=p_y\in\{0,\frac{1}{2}\}$. This defines a $\mathbb{Z}_2$ index. In the context of charge neutral systems, the polarization of Eq.~\eqref{eqn:polarization} should be understood as the expectation value of the dipole moment: ${\bf P}=\langle \Psi|{\bf r}|\Psi \rangle$. The irreducible representations at HSPs in the first BZ can be used to determine the system's polarization. In particular (\italicize{54}):
\begin{linenomath}
\begin{equation}
        \label{eqn:polarizationBenalcazar}
      {\bf P}=\frac{1}{2}\left[X_1-\Gamma_1\right]({\bf \hat{x}}+{\bf \hat{y}}) \quad \text{mod} \,1,
\end{equation}
\end{linenomath}
where $K_\ell$ is the number of bands below the gap that transform according to the representation $\ell$ at the high symmetry point $K$. Applying Eq.~\eqref{eqn:polarizationBenalcazar} to our model, we obtain $\left(p_x,p_y\right) = \left(1/2,1/2\right)$.

A similar formula exists for corner modes (\italicize{54}). For $C_4$--symmetric systems, the quantized corner charges are given by
\begin{linenomath}
\begin{equation}
        \label{eqn:cornerBenalcazar}
      Q=\frac{1}{4}\left[X_1+2M_1+3M_2-3\Gamma_1-3\Gamma_2\right]\quad \text{mod} \,1.
\end{equation}
\end{linenomath}
Our system has $Q=1/4$. Eq.~\eqref{eqn:cornerBenalcazar} defines a $\mathbb{Z}_4$ topological index for $C_4$--symmetric system. Nevertheless, this $\mathbb{Z}_4$ index is well defined only for a trivial $\mathbb{Z}_2$ index associated to the polarization of the system.
In fact, a non-zero bulk polarization can manifest itself with in-gap dispersive modes at the edges of the system. Corner modes can mix with the extensive number of dispersive states and cannot be individually resolved. 

\begin{figure}
	\begin{center}
		\includegraphics{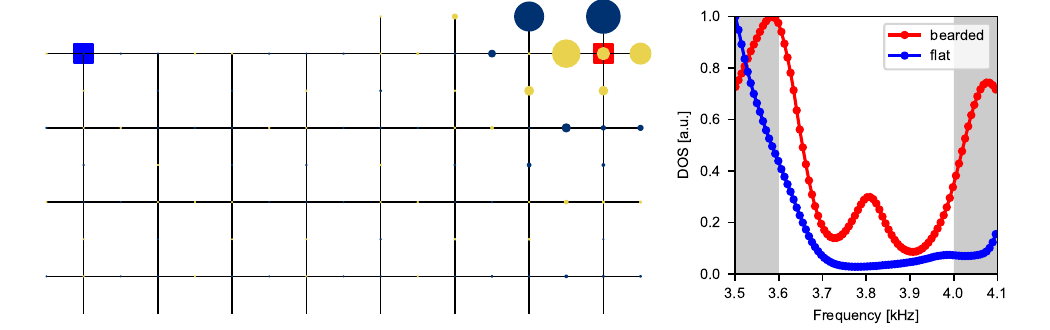}
	\end{center}
	\caption{{\bf Corner modes.} In the left panel we show the simulated mode profile of the  corner mid-gap state in Fig.~\ref{fig:SI-FlowingSates}.A. On the left side a flat edge, and on the right side a bearded one ending with $2c$ states. The size of the circles indicate the weight, the color sign structure of the mode. The two squares (red/blue) denote the sites where we measured the experimental response. In the right panel we show the measured response curves at the two $A$--sites closest to the cut. One can see how the bearded edge nucleates a corner mode, while the flat edge does not host any in-gap state.}
	\label{fig:SI-CornerMode}
\end{figure}

The presence of a quantized polarization in our system renders the corner modes ill-defined. As discussed in Sec.~\ref{section:SCNstates}, the edge modes of fragile topological insulators are not robust and can be removed from the gap by the application of symmetry preserving perturbations, although a filling anomaly remains. Such perturbations have been applied to remove the edge modes from the bulk gap of our model and observe the spectral signatures of fragility. Possible mid-gap corner modes have also been removed as finite-elements simulations and measurements show no evidence of modes strongly localized at the corners of the sample. 
Nonetheless, simulations for a system with a full $C_2$--symmetric cut ($\lambda=0$) display strongly localized modes at corners. Along the cut, two different terminations coexist. One one hand, there is a side ending with $1b$ sites and realizing a flat edge. On the other hand, the opposite side realizes a bearded edge that terminates with $2c$ sites, cf. Fig.~\ref{fig:SI-CornerMode}.A. The corner modes are strongly localized close to the cut along the bearded edges, as can be seen in Fig.~\ref{fig:SI-CornerMode}.A. In Fig.~\ref{fig:SI-CornerMode}.B, we present the measured spectral response around the gap frequency at the corner for the bearded and the flat terminations

These observations suggest that the presence of spectrally isolated corner modes in fragile topological insulators depends on the details of the boundary terminations. 
It would be interesting to study whether recent proposals to detect bound states in the continuum of higher order topological insulators could be applied to detect corner states in fragile topological insulators (\italicize{55}). 

\subsection{Diagnosis of fragile topology via inversion symmetry eigenvalues}

A diagnosis of fragile topology based on inversion symmetry eigenvalues at HSPs has been proposed in (\italicize{19}) for inversion symmetric systems. Here, we verify that our model realizes a fragile topological insulator also according to this classification scheme. 

One looks at the irreducible representation at inversion symmetric points in the first Brillouin zone: $\Gamma=\left(0\,,0\right)$, $X=\left(\pi\,,0\right)$, $Y=\left(0\,,\pi\right)$, $M=\left(\pi\,,\pi\right)$. The diagnosis of fragile topology is based on the vector:
\begin{linenomath}
\begin{equation}
        \label{eqn:fragileInversion}
      v_{\bf k}=\left(n,\nu_{\Gamma},\nu_{X},\nu_{Y},\nu_{M} \right),
\end{equation}
\end{linenomath}
where $n$ is the number of bands below the energetic gap and $\nu_{K}$ is the difference between the number of odd and even irreducible representation under inversion at the HSP $K$. For our system, it is straightforward to compute $v_{\bf k}=\left(2,-2,2,2,2\right)$. This vector contains the inversion symmetry information in momentum space. A simple matrix introduced in (\italicize{19}) allows to map this reciprocal space information to real space data: $v_{\bf{r}}=A\,v_{\bf{k}}$ , where:
\begin{linenomath}
\begin{equation}
        \label{eqn:fragileInversionMatrix}
      A=\frac{1}{4}\begin{pmatrix}
      \phantom{-}4 & \phantom{-}0 &\phantom{-} 0 & \phantom{-}0 & \phantom{-}0 \\
      \phantom{-}0 & \phantom{-}1 & \phantom{-}1 & \phantom{-}1 & \phantom{-}1 \\
      \phantom{-}0 & \phantom{-}1 & -1 & \phantom{-}1 & -1 \\
      \phantom{-}0 & \phantom{-}1 & \phantom{-}1 & -1 & -1 \\
      \phantom{-}0 & \phantom{-}1 & -1 & -1 & \phantom{-}1 
      \end{pmatrix}\,.
\end{equation}
\end{linenomath}
For the case at hand, we obtain $v_{\bf r}=\left(2, 1, 0, -1, -1\right)$. Whenever all 
coefficient of $v_{\bf r}$ are integers, the insulator cannot be a stable topological 
insulator. On the other hand, fragile topology is realized if (\italicize{19})
\begin{linenomath}
\begin{equation}
        \label{eqn:fragileInversionCriteria}
      \sum_{i=1}^4 |v_{\bf r}^i|>v_{\bf r}^0\,.
\end{equation}
\end{linenomath}
Applying criterion~\eqref{eqn:fragileInversionCriteria} to our model, we establish the fragile topology entirely from the inversion symmetry eigenvalues at HSP in the first BZ.

\subsection{Wilson loop and fragile topology}
Non-trivial topology can be captured by the Berry phases of the bands under consideration. Non-Abelian Wilson loops are a generalization of Berry phase for multi-band systems (\italicize{56}--\italicize{59}). Given that our model has two non-separable bands ($N=2$) below the gap of interest, we might try to capture the non-trivial topology of the system with them.
The Bloch Hamiltonian $H({\bf k})$ is parametrized by ${\bf k}$ in the first BZ. The non-Abelian Berry curvature is defined as:
\begin{linenomath}
  \begin{equation}
    A_{m,n}({\bf k})=\langle u_{{\bf k},m} |\nabla_{\bf k}| u_{{\bf k},n}\rangle, \quad n.m\in\{1,...,N\},
  \end{equation}
\end{linenomath}
where $\ket{u_{{\bf k},j}}$ is the Bloch wavefunction at momentum ${\bf k}$ for the $j^{\,\text{th}}$ band. The Wilson loop is then defined as:
\begin{linenomath}
  \begin{equation} 
    \label{eqn:Wilson}   W\left[\mathcal{C}\right]=\mathcal{P}\,\text{exp}\left[\oint_\mathcal{C}\text{d}\ell\cdot A({\bf k})\right],
  \end{equation}
\end{linenomath}
where $\mathcal{C}$ is a closed path in momentum space and $\mathcal{P}$ indicates path ordering such that operators at the beginning of the path appear to the right of operators at the end. $WW^\dagger=\mathbb{I}$ and the eigenvalues of $W\left[\mathcal{C}\right]$ are gauge invariant complex numbers of unit modulus: $\text{exp}\,[i\theta]$. The phases of these eigenvalues (Wilson spectrum) are related to the spectrum of the position operator projected on the space of bands below the gap (\italicize{60}). The eigenvalues of $W$ along any non-contractible loop $\mathcal{C}$ defines a map $S^1\rightarrow S^1$. In the presence of crystalline symmetries, their winding number can serve as topological index.

Wilson loops can be computed efficiently. One considers a discretized version of Eq.~\eqref{eqn:Wilson}:
\begin{linenomath}
  \begin{equation} 
    \label{eqn:WilsonDisc}  
    W\left[\mathcal{C}\right]=F_{{\bf k}_0,{\bf k}_1}F_{{\bf k}_1,{\bf k}_2}\dots F_{{\bf k}_{p-1},{\bf k}_p},
  \end{equation}
\end{linenomath}
where ${\bf k}_0={\bf k}_p$ and $F_{{\bf k}_i,{\bf k}_j}^{m,n}=\langle u_{{\bf k}_i,m} | u_{{\bf k}_j,n}\rangle$. From this formulation, it is clear how Wilson loops can be regarded as matrices that map a starting point ${\bf k}_0$ in momentum space to another point along the loop under parallel transport. Extra care needs to be taken when computing Eq.~\eqref{eqn:WilsonDisc}, as discretization can introduce non-unitary effects, which vanish in the limit of large number of points. One should perform a singular value decomposition $W=UDV^\dagger$ and consider instead the matrix:
\begin{linenomath}
  \begin{equation} 
    W'=UV^\dagger,
  \end{equation}
\end{linenomath}  
which is unitary. All the Wilson loops presented in this work have been calculated with the Z2Pack library (\italicize{61}, \italicize{62}).

The presence of crystalline symmetries imposes constraints on the Wilson loop spectrum. In particular, we are interested in what happens for inversion and time-reversal symmetric insulators in two spatial dimensions (\italicize{60}). In 2D, the Wilson loops along one direction, e.g., ${\bf \hat{x}}$, are functions of the initial momentum along the other direction: $W_{k_y}$. Inversion symmetry constraints:
\begin{linenomath}
  \begin{equation}
    \text{exp}\,[i\theta_{k_y}]=\text{exp}\,[-i\theta_{-k_y}].
  \end{equation}
\end{linenomath}
Time-reversal symmetry, on the other hand, requires that:
\begin{linenomath}
  \begin{equation}
    \text{exp}\,[i\theta_{k_y}]=\text{exp}\,[i\theta_{-k_y}].
  \end{equation}
\end{linenomath}
This has important consequences at HSPs $k_y=\{0,\pi\}$ where the Wilson loop eigenvalues have to be degenerate at $\pm 1$ or come in complex conjugate pairs.

It turns out the the winding of the Wilson loops can be entirely determined by looking at the inversion eigenvalues of the Bloch bands at HSPs (\italicize{60}). Fixing $K_x=\{0,\pi\}$, one defines the number of even and odd bands under inversion symmetry at ${\bf k}_{\text{inv}}=\{(K_x,0),(K_x,\pi)\}$ as $n_+({\bf k}_{\text{inv}})$ and $n_-({\bf k}_{\text{inv}})$, respectively and considers the vector $\{n_+(k_y=0),n_-(k_y=0),n_+(k_y=\pi),n_-(k_y=\pi)\}$. The smallest component of the vector is indicated as $n_s$, the $k_y$ at which it is realized as $k_s$ and the inversion eigenvalue as $s$. Provided these quantities one can then map the inversion symmetry eigenvalues to the Wilson loop spectrum which has:
\begin{enumerate}
  \item $\left[n_+(k_y=k_s+\pi)-n_s\right]$ number of $-s$ eigenvalues,
  \item $\left[n_-(k_y=k_s+\pi)-n_s\right]$ number of $s$ eigenvalues,
  \item $n_s$ pairs of complex conjugate eigenvalues.
\end{enumerate}
Further discussions and explanations of this mapping are provided in (\italicize{60}).

\begin{figure}
	\begin{center}
		\includegraphics{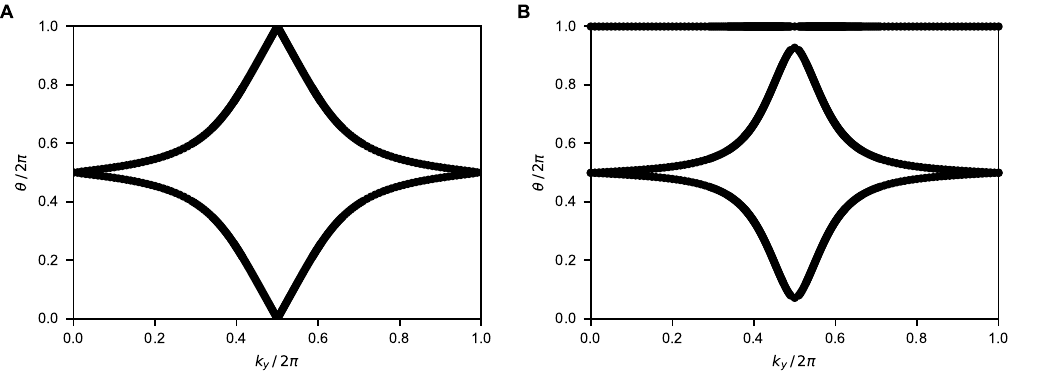}
	\end{center}
	\caption{{\bf Wilson loop spectrum.} {\bf A} For the simple tight binding model of Eq.~\eqref{eqn:tbModel}. {\bf B} With the addition of a $d_{xy}$ orbital at the $1a$ maximal Wyckoff position.}
	\label{fig:SI-WilsonLoop}
\end{figure}

Here, we apply the mapping to the concrete case of the lower two bands of the tight binding model of Eq.~\eqref{eqn:tbModel}. Starting from the irreducible representations of Table~\ref{tab:TBirreducible representations}, we can perform the mapping as shown in Table~\ref{tab:WLMap}. The Wilson loop spectrum obtained by the inversion symmetry eigenvalues confirms the winding of the Wilson loop. This observation is validated by direct calculations as shown in Fig.~\ref{fig:SI-WilsonLoop}.A and proves the non-trivial topology of these bands. 
\begin{table}[!ht]
  \begin{tabular}{ccccccc}
    \hline
    \hline
    \multicolumn{4}{c}{$c_2$ eigenvalues}  & {\phantom{(0)}} & \multicolumn{2}{c}{$W_{k_y}$ eigenvalues}\\
  $(0,0)$ & $(0,\pi)$ & $(\pi,0)$ & $(\pi,\pi)$ & {\phantom{(0)}} & $k_y=0$ & $k_y=\pi$\\
  \cline{1-4}
   \cline{6-7}
  $(+,+)$ & $(-,-)$ & $(-,-)$ & $(-,-)$  & {} & $[-,-]$ & $[+,+]$ \\
  \hline
  \hline
  \end{tabular}
  \caption{{\bf Mapping from inversion eigenvalues to Wilson spectrum.} The relation between inversion eigenvalues at different momentum points (as described in the second row) and the Wilson spectrum of the studied system. The pair of opposite eigenvalues of the Wilson loop at $k_y=0$ and $k_y=\pi$ ensure spectral flow.}
  \label{tab:WLMap}
  \end{table}

\subsection{Trivialization of fragile bands by the addition of trivial bands}
The main feature of fragile topological insulators is that they can be turned trivial by the addition of topologically trivial bands below the energetic gap of interest. The lower two bands of our model realize the EBR decomposition:
\begin{linenomath}
\begin{equation}
(A_1)_{1b}\oplus (A_2)_{2c} \ominus (B_2)_{1a}\;.
\end{equation}
\end{linenomath}
The addition of a single EBR induced by a $d_{xy}$ orbital ($B_2$) at Wyckoff position $1a$ suffices to realize a trivial atomic insulator.
In the following, we confirm this observation in a simple model. To capture the additional orbital, at least theoretically, one could consider the tight binding Hamiltonian:
\begin{linenomath}
\begin{equation}
  \label{eqn:TBTrivial}
  H(k_x\,,k_y)=\begin{pmatrix}
  2t_0 & t_d & 0 & 0 & t_d & 0 & 0  \\
  t_d^* & m & t_x & t_y  & 0 & 0 & 0 \\
  0 & t_x^* & 0 & 0  & 0 & 0 & t_{-} \\
  0 & t_y^* & 0 & 0  & 0 & t_{+} & 0 \\
  t_d^* & 0 & 0 & 0 &  m & t_x & t_y \\
  0 & 0 & 0 & t_{+}^* & t_x^* & 0 & 0 \\
  0 & 0 & t_{-}^* & 0 &  t_y^* & 0 & 0 \\
  \end{pmatrix},
\end{equation}
\end{linenomath}
where $t_{x,y}=t_0\left(1+e^{+ik_{x,y}}\right)$, $t_{-}=\chi\left(1+e^{-i(k_x -k_y)}\right)$, $t_{+}=\chi\left(e^{ik_x}+e^{-ik_y}\right)$ and $t_d=2\chi\left(1+e^{-i(k_x+k_y)}-e^{-ik_x}-e^{-ik_y}\right)$. Note that the $d_{xy}$ nature of the added orbital is encoded in the non-trivial sign structure of its couplings to the $1b$ sites. The choice of parameters has been made such that the additional band spectrally overlaps with the two lower bands of the original model.

A first indication of the trivial nature of the extended model comes from the Wilson loop spectrum for this system. In fact, it does not show spectral flow after the addition of the $d_{xy}$ orbital, as can be seen in Fig.~\ref{fig:SI-WilsonLoop}.B. A final confirmation is given by the analysis of the irreducible representations at HSPs that confirms the EBR decomposition of an atomic insulator:
\begin{linenomath}
\begin{equation}
(A_1)_{1b}\oplus (A_2)_{2c}\;.
\end{equation}
\end{linenomath}

\begin{figure}
	\begin{center}
		\includegraphics[width=2.25in]{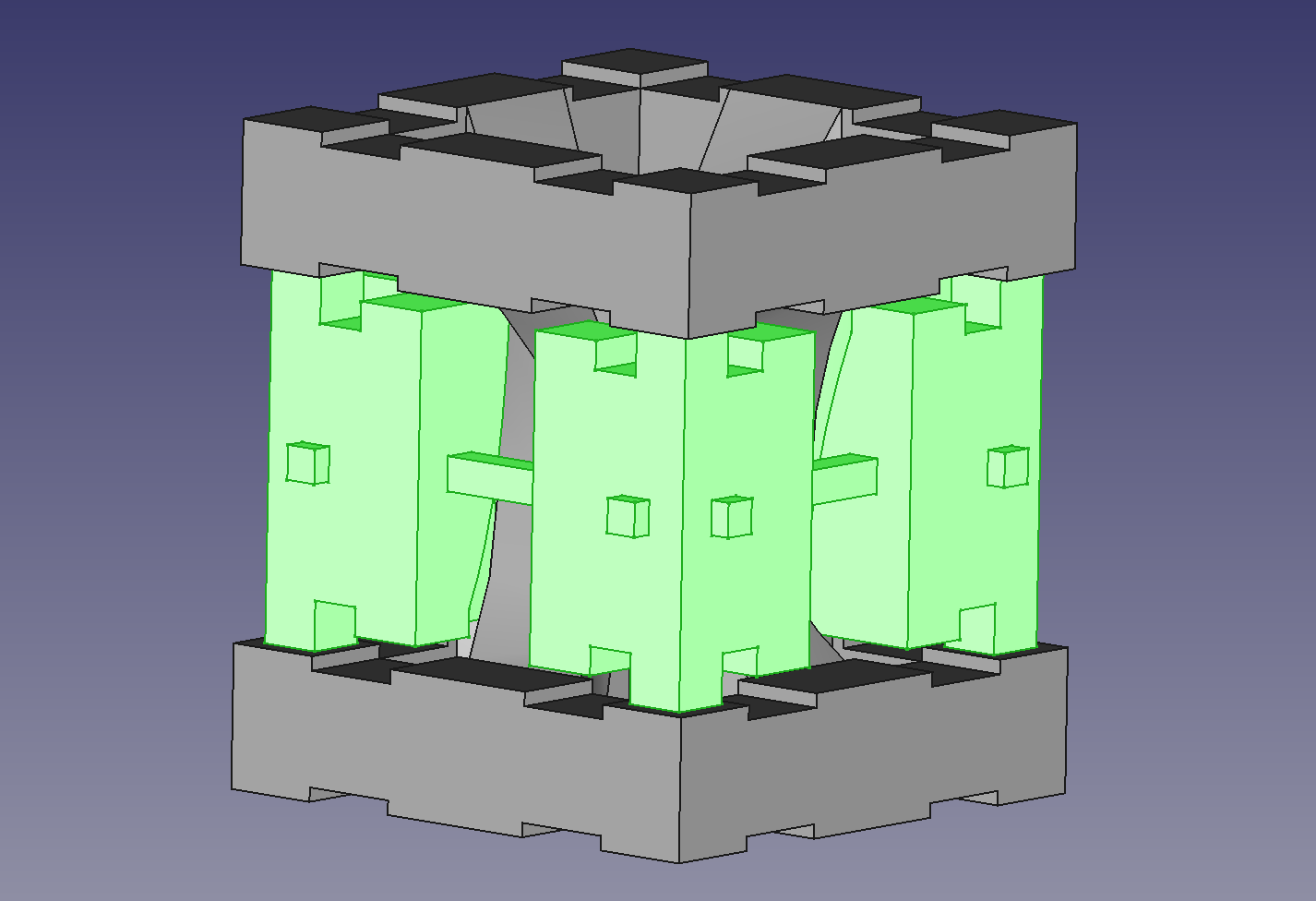}
	\end{center}
	\caption{{\bf Illustration of trivializing bands.} Rendering of a simple unit cell that allows to realize a mode that transforms as a $d_{xy}$ orbital at low frequencies in our sonic crystal.}
	\label{fig:SI-Trivialize}
\end{figure}

One might wonder how feasible the proposed trivialization scheme is in an acoustic metamaterial. In fact, modes with a $d_{xy}$ symmetry are usually high in frequency and will not overlap with the lowest modes of the cavities considered in our sample. It is however possible to realize an effective $d_{xy}$ mode by coupling four individual cavities at non-maximal Wyckoff positions around the site $1a$, cf. Fig.~\ref{fig:SI-Trivialize}. These four resonators will give rise to four extra bands. A dimerization pattern of the intra-unit cell and inter-unit cell hoppings opens gaps among them. The four bands realize the EBR decomposition:
\begin{linenomath}
\begin{equation}
(A_1)_{1a}\oplus (B_2)_{1a} \oplus (E)_{1a}\;.
\end{equation}
\end{linenomath}
This construction provides a practical way to obtain a mode that transforms as a $d_{xy}$ orbital at low frequencies. One can then tune the size of the cavities and the diameter of the tubes mediating the hopping to have the relevant orbital spectrally overlapping with the modes of interest.

\subsection{Fragile topology in other classical systems}
Most of the reported topological bands in classical wave systems can be interpreted as fragile crystalline systems (\italicize{18}, \italicize{32}). The array of coupled pendula (\italicize{63}) by some of the present authors are no exception. The relevant matrix describing the dynamics of the pendula of (\italicize{63}) is given by 
\begin{equation}
	\label{eqn:SI-DynamicalPendula}
\mathcal D=-\begin{pmatrix}
	\Delta -\cos(k_y) & 1 & e^{-ik_x}&i\sqrt{3}\sin(k_y) & 0 &0 \\
	1 & 2\cos(k_y) & 1 &0 &0&0\\
	e^{ik_x} & 1 & \Delta-\cos(k_y) & 0 &0 &-i\sqrt{3} \sin(k_y)\\ 
	-i\sqrt{3} \sin(k_y)&0&0&-\cos(k_y) & 1 & e^{-ik_x} \\
	0&0&0&1 & 2\cos(k_y) & 1 \\
	0 &0& i\sqrt{3} \sin(k_y)&e^{ik_x} & 1 & -\cos(k_y)\\
\end{pmatrix}.
\end{equation}
The $6\times 6$ structure stems from the fact that we have a unit cell with three sites in $x$--direction with two pendula (top-left and bottom-right block) per site. This corresponds to a Landau gauge for two copies of a flux $\pm 1/3$ Hofstadter model. For $\Delta=0$, $\mathcal D$ can be brought to a block-diagonal form using 
\begin{equation}
	S=\frac{1}{\sqrt{2}}\begin{pmatrix} 1 & i \\ 1 & -i \end{pmatrix}.
\end{equation}
This corresponds to a local (per site) symmetry of 
\begin{equation}
	U = \openone_{3\times 3}\otimes i\sigma_y,
\end{equation}
which reflects that both local pendula see the same environment.

This setup allows for a ``strong'' topological index, if one accepts the necessity of the engineered symmetry $U$: One can go to the eigen-sectors of $U$, where the dynamical system is identical to a Hofstadter problem. Therefore, the system is characterized by Chern numbers, albeit of different sign in each sector. Or, one realizes that classical time reversal acts simply as complex conjugation $K$ and hence can be combined with $U$ to 
\begin{equation}
	\mathcal T =\openone_{3\times 3}\otimes i\sigma_y \circ K,
\end{equation}
which is nothing but the standard fermionic time reversal with $\mathcal T^2=-\openone$. However, note that here $K$ and $U$ are individual symmetries and hence the Chern number per $U$--sector might be a more accurate topological index than the $\mathbb Z_2$ index of symmetry class AII.

Given that in each $U$--sector we have a Chern number, the bands of $\mathcal D$ are non-Wannierizable. On top of that, we formulated the dynamical matrix of the pendula $\mathcal D$ in momentum space, and hence, we deal with a crystalline system. This begs the question of what type this non-Wannierizability is from a crystalline point of view: stable or fragile. To answer this question, we break the ``strong'' Chern number by breaking the $U$ symmetry via $\Delta \neq 0$. By doing so, we remain with the wallpaper group $p2$.
\begin{figure}[t]
	\begin{center}
		\includegraphics{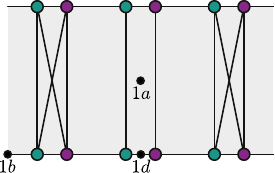}
	\end{center}
	\caption{{\bf Sketch of the pendula setup.}(\italicize{63}) Colored dots are the location of the pendula, black lines indicate couplings and the grey area is the unit cell. The maximal Wyckoff positions of the wallpaper group $p2$ are shown.}
	\label{fig:SI-pendula}
\end{figure}
\begin{figure}[t]
	\begin{center}
		\includegraphics{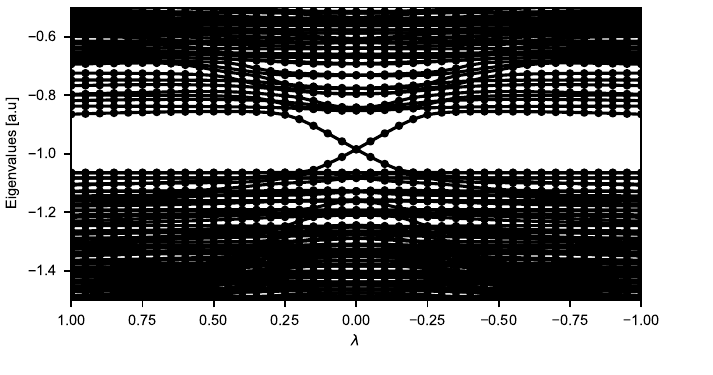}
	\end{center}
	\caption{{\bf Spectral flow for soft cut in coupled pendula.} (\italicize{63}) Eigenstates of the matrix (\ref{eqn:SI-DynamicalPendula}) for a soft cut described by the parameter $\lambda$ which multiplies all hoppings over the $C_2$ symmetric cut.}
	\label{fig:SI-FlowPendula}
\end{figure}

We analyze the symmetry properties of $\mathcal D$, see Fig.~\ref{fig:SI-pendula} for a sketch of the couplings and the definition of the maximal Wyckoff positions. For $p2$, we only need the matrix for two-fold rotations around the $1a$ Wyckoff position, which in our model is realized as 
\begin{equation}
	c_2 = 
	\begin{pmatrix}
	0 & 0 & e^{ik_y} & 0 &0 & 0 \\ 
	0 & e^{ik_y} & 0 & 0 &0 & 0 \\ 
	e^{ik_y} & 0 & 0 & 0 &0 & 0 \\ 
	0 & 0 & 0 & 0 &0 & e^{ik_y} \\ 
	0 & 0 & 0 & 0 &e^{ik_y} & 0 \\ 
	0 & 0 & 0 & e^{ik_y} &0 & 0 
	\end{pmatrix}.
\end{equation}
Evaluated for the set of six bands and compared with the possible EBRs (\italicize{42}), we conclude that the system of pendula is described by the following decomposition into EBRs
\begin{align}
	\text{bands 1\&2} &: (A)_{1b} \oplus (B)_{1c} \oplus (B)_{1d} \ominus (A)_{1a}, \label{pendulaMiss}\\
	\text{bands 3\&4}  &: (A)_{1a} \oplus (B)_{1a} \oplus (A)_{1d} \oplus (B)_{1d}\ominus 2(B)_{1c},\\
	\text{bands 5\&6}  &: (A)_{1b} \oplus (B)_{1c} \oplus (A)_{1d} \ominus (B)_{1a}. 
\end{align}

With the above we establish the fragility of the crystalline topology at $\Delta\neq 0$. This is a rather generic feature: without spin-orbit coupling and with spinless time reversal symmetry all non-Wannierizable bands are of fragile nature (\italicize{42}). 

As a sanity check we perform a soft cut with $C_2$--symmetry. In particular, the symmetry center of the cut goes through the $1a$ Wyckoff position, from which the EBR with the negative coefficient is induced in \eqref{pendulaMiss}. In Fig.~\ref{fig:SI-FlowPendula} we show the resulting spectral flow, compatible with $\delta=1$, cf. Sec.~\ref{section:twisted}.